\documentclass[aps,prx,twocolumn,superscriptaddress]{revtex4-1}

\usepackage{graphicx}
\usepackage{subfigure}
\usepackage{amsmath}
\usepackage{amssymb}
\usepackage{mathrsfs}
\usepackage{hyperref}
\usepackage[utf8]{inputenc}
\usepackage{mathtools}
\usepackage[english]{babel}
\hypersetup{colorlinks=true, linkcolor=black, citecolor=blue, urlcolor=blue}

\newcommand{\ket}[1]{\left| #1 \right>} 
\newcommand{\bra}[1]{\left< #1 \right|} 

\newcommand{\ie}{i.\,e.\ }

\newcommand{\fref}[1]{\text{Fig.}~\ref{#1}}
\newcommand{\eref}[1]{\text{Eq.}~\eqref{#1}}

\begin{document}
\title{Cavity-induced emergent topological spin textures in a Bose Einstein condensate}
\author{Stefan Ostermann}
\email[Corresponding author: ]{stefan.ostermann@uibk.ac.at}
\affiliation{Institut f\"ur Theoretische Physik, Universit\"at Innsbruck, Technikerstraße 21a, A-6020 Innsbruck, Austria}
\author{Hon-Wai Lau}
\affiliation{Max-Planck-Institut f\"{u}r Physik komplexer Systeme, D-01187 Dresden, Germany}
\affiliation{Institute for Quantum Science and Technology and Department of Physics and Astronomy, University of Calgary, Calgary T2N 1N4, Alberta, Canada}
\author{Helmut Ritsch}
\affiliation{Institut f\"ur Theoretische Physik, Universit\"at Innsbruck, Technikerstraße 21a, A-6020 Innsbruck, Austria}
\author{Farokh Mivehvar}
\email[Corresponding author: ]{farokh.mivehvar@uibk.ac.at}
\affiliation{Institut f\"ur Theoretische Physik, Universit\"at Innsbruck, Technikerstraße 21a, A-6020 Innsbruck, Austria}

\begin{abstract}
The coupled nonlinear dynamics of ultracold quantum matter and electromagnetic field modes in an optical resonator exhibits a wealth of intriguing collective phenomena. Here we study a $\Lambda$-type, three-component Bose-Einstein condensate coupled to four dynamical running-wave modes of a ring cavity, where only two of the modes are externally pumped. However, the unpumped modes play a crucial role in the dynamics of the system due to coherent back-scattering of photons. On a mean- field level we identify three fundamentally different steady-state phases with distinct characteristics in the density and spatial spin textures: a combined density and spin wave, a continuous spin spiral with a homogeneous density, and a spin spiral with a modulated density. The spin-spiral states, which are topological, are intimately related to cavity-induced spin-orbit coupling emerging beyond a critical pump power. The topologically trivial density-wave--spin-wave state has the characteristics of a supersolid with two broken continuous symmetries. The transitions between different phases are either simultaneously topological and first order, or second order. The proposed setup allows the simulation of intriguing many-body quantum phenomena by solely tuning the pump amplitudes and frequencies, with the cavity output fields serving as a built-in nondestructive observation tool.
\end{abstract}

\maketitle

\section{Introduction}
\label{sec:introduction}

The experimental progresses in reaching the quantum degeneracy limit in atomic gases paved the way for the realization of quantum many-body phenomena in these highly tunable systems~\cite{Lewenstein-2007,Bloch-2008}. Some of the most remarkable examples include the realization of the superfluid to Mott-insulator quantum phase transition~\cite{Greiner-2002}, quantum magnetism and magnetic orderings~\cite{Struck2013,Greif2013,Greif2015,Hart2015,Mazurenko2017}, synthetic magnetic fields (i.e., Abelian gauge potentials)~\cite{Lin2009,Aidelsburger2013,Miyake2013}, and spin-orbit coupling (i.e., non-Abelian gauge potentails)~\cite{lin2011spin,Wu2016,Li2017stripe} in ultracold quantum gases. While the first generation of experiments was limited to static lattices and local contact interactions, the study of highly nonlinear optical systems, where the back-action of the matter on the radiation fields is not negligible, has opened up new frontiers towards dynamical optical potentials, long-range atom-atom interactions, and exotic collective phenomena~\cite{Ritsch-2013,ostermann2016spontaneous,dimitrova2017observation,vochezer2018light}. The most prominent examples include the coupling of ultracold atoms to dynamic quantized radiation fields of high-quality cavities~\cite{Kruse2003,Nagorny2003,Brennecke-2007}, leading to the realization of the Dicke superradiance phase transition~\cite{Baumann-2010,Klinder2015}, atomic recoil lasing ~\cite{Kruse-2003,Slama-2007,Schmidt-2014}, and the quantum phase transition between superfluid, supersolid, Mott-insulator, and density-wave phases~\cite{Landig-2016,leonard2017supersolid}.

\begin{figure}[b!]
\centering
\includegraphics[width=0.44\textwidth]{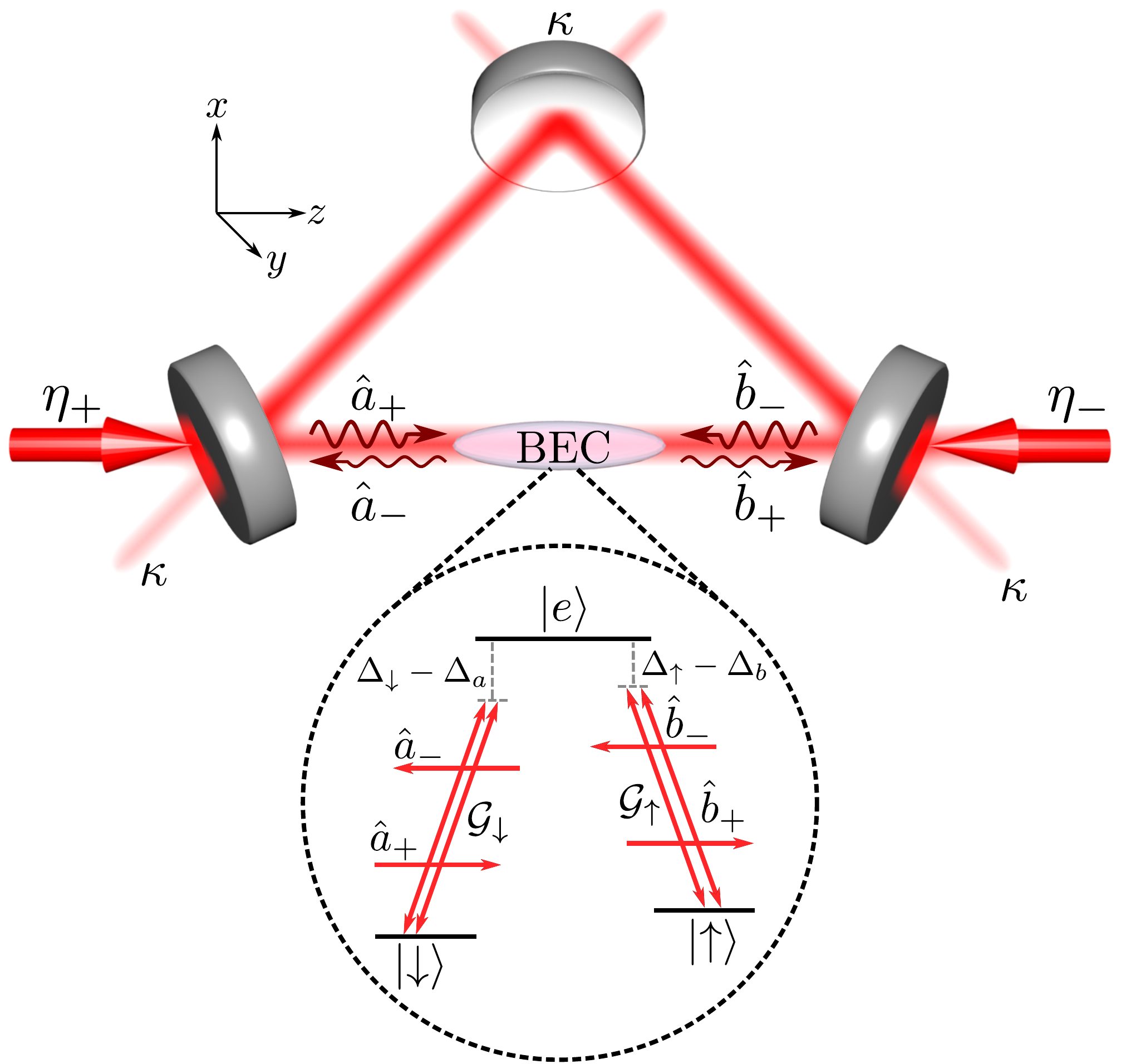}
\caption{Sketch of the system. A spin-1 BEC is tightly confined along one leg of a longitudinally pumped ring cavity with four nearly resonant modes $\{\hat{a}_\pm e^{\pm ikz},\hat{b}_\pm e^{\pm ikz}\}$. The modes $\hat{a}_\pm$ ($\hat{b}_\pm$) induce the transitions $\ket{\downarrow}\leftrightarrow\ket{e}$ ($\ket{\uparrow}\leftrightarrow\ket{e}$) with a coupling strength $\mathcal{G}_{\downarrow}$ ($\mathcal{G}_{\uparrow}$). The modes $\hat{a}_+$ and $\hat{b}_-$ are pumped by external lasers with strengths $\eta_+$ and $\eta_-$, respectively.}
\label{fig:setup}
\end{figure}

Almost all experimental works and most theoretical studies of coupled atom-cavity systems in the past were limited to systems where either the atomic internal states~\cite{Dimer2007,Larson2009,Gopalakrishnan2011,Zhiqiang2017} or the atomic external degrees of freedom~\cite{Moore1999,Domokos2002,nagy2006self,Maschler2008,Nagy2008,nagy2010dicke,Piazza2013,Piazza2014,Keeling2014,Piazza2015, Mivehvar2017a,Mivehvar2018} are taken into account. Only recently theoretical investigations towards including both atomic internal and external degrees of freedom in cavity QED have been conducted. These systems exhibit many more interesting phenomena, including the emergence of synthetic strong magnetic fields and spin-orbit coupling~\cite{Mivehvar-2014,Dong-2014,Deng-2014,Mivehvar-2015}, disorder-driven density and spin self-ordering~\cite{Mivehvar2017b}, topological states~\cite{Pan2015,Yu2017}, and a variety of magnetic orders~\cite{Fan-2017,colella2018quantum}. Very recently, the first  experimental implementation with a spin-1 BEC inside a linear cavity revealed spontaneous self-ordering of the atoms into a crystalline structure with an antiferromagnetic order~\cite{landini2018formation}.

In this work we study a novel type of hybrid atom-cavity system, where $\Lambda$-type spin-1 ultracold bosons are confined into quasi one dimension along one leg of a ring cavity with two ``pairs'' of nearly resonant modes~\cite{gangl2001cavity,gangl2002cavity}. Each pair consists of two counterpropagating modes with the same polarization which is orthogonal to the polarization of the other pair. In contrast to Ref.~\cite{ostermann2015atomic}, each pair of modes only couples to one of the transitions in the $\Lambda$ scheme, where the atoms are assumed to posses two ground states (e.g., two different Zeeman sublevels) and an electronic excited state; see Fig.~\ref{fig:setup}. The adiabatic elimination of the upper atomic electronic state as well as of other Zeeman sublevels results in an effective two-component pseudospin model. We consider a case where two counterpropagating modes of orthogonal polarizations are externally pumped by lasers through the cavity mirrors as depicted in Fig.~\ref{fig:setup}. These two modes do not interfere with one another and the system is, therefore, initially homogeneous. This is reminiscent of the scheme for generating equal Rashba-Dresselhaus synthetic spin-orbit coupled for neutral atoms in free space~\cite{lin2011spin}.

That said, in the present case the cavity modes are dynamic and affected by the atomic dynamics as well as by photon losses through the cavity mirrors. Crucially, the unpumped modes are ``dynamically'' populated by coherently scattered photons and, therefore, couple to the pumped modes and the atomic internal and external degrees of freedom. As an important consequence, the cavity-induced spin-orbit coupling for the atoms only emerges above a critical pump power, which in turn gives rise to novel nonequilibrium quantum phases and quantum phase transitions of various natures in our system. It is this dynamical population of the unpumped cavity modes and its nontrivial interplay with the other degrees of freedom which marks a sharp contrast to the free-space spin-orbit coupling~\cite{lin2011spin,Wu2016,Li2017stripe,Li2012,Li2013} as well as all other previous cavity-based spin-orbit coupling schemes~\cite{Mivehvar-2014,Dong-2014,Deng-2014,Pan2015}.

\begin{figure}[t]
\centering
\includegraphics[width=0.48\textwidth]{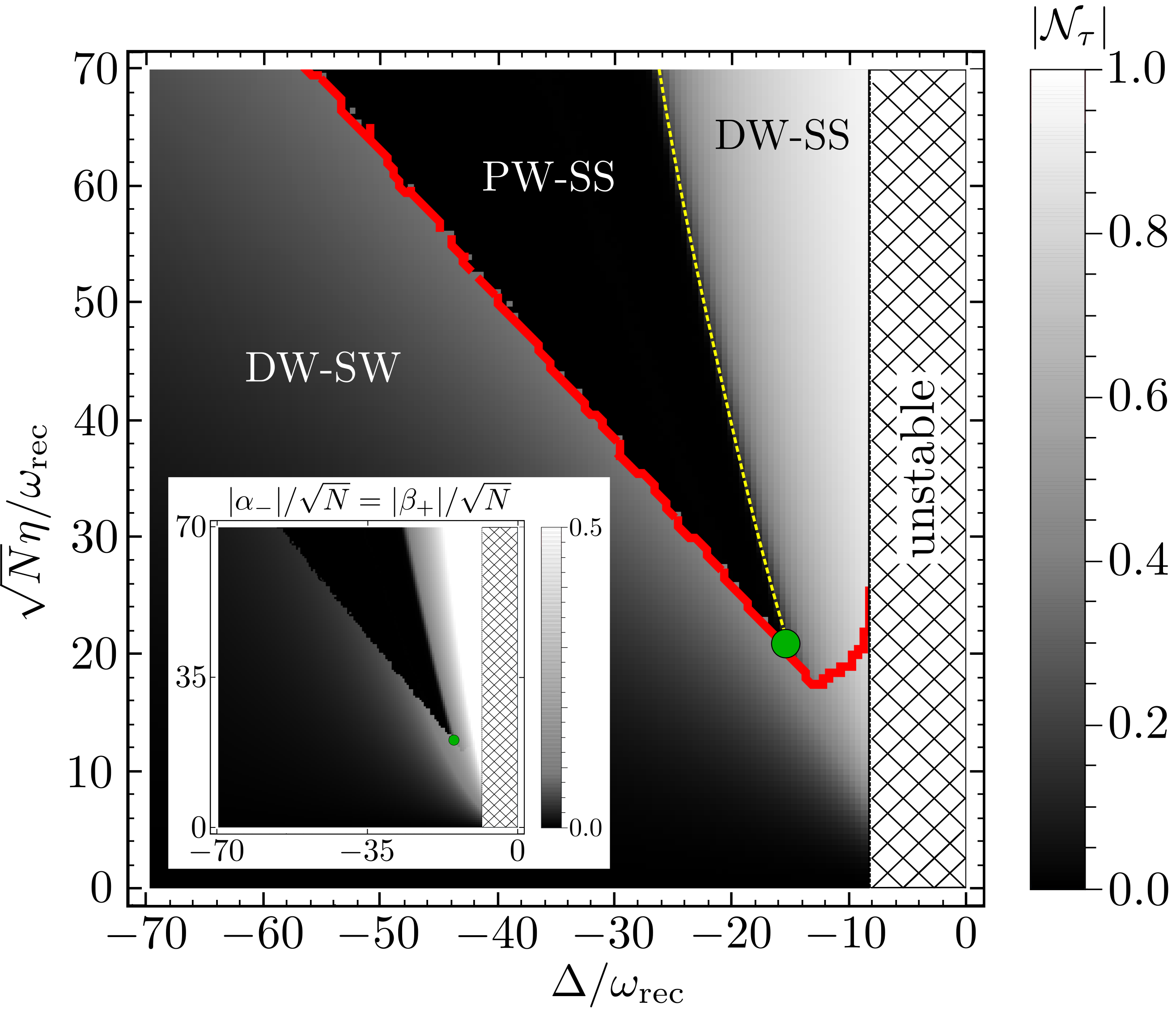}
\caption{The mean-field atomic phase diagram in the rescaled parameter space $\{\sqrt{N}\eta/\omega_{\rm rec},\Delta/\omega_{\rm rec}\}$. It displays three distinct phases: density-wave--spin-wave (DW-SW), plane-wave--spin-spiral (PW-SS), and density-wave--spin-spiral (DW-SS) states. The color coding shows the absolute values of the density-wave order parameters, $|\mathcal{N}_\downarrow|=|\mathcal{N}_\uparrow|$. The solid red curve marks the onset of the topological phase transition. The dashed yellow curve indicates the boundary of the second-order phase transition and the green dot indicates the tricritical point. The unstable parameter regions are marked by the hased pattern. The inset illustrates the cavity-field phase diagram in the same parameter space, where the color coding shows the absolute values of the cavity-field order parameters, $|\alpha_-|\sqrt{N}=|\beta_+|/\sqrt{N}$. It coincides precisely with the atomic phase diagram. The other parameters are set to $(U_0,\Omega_{0 \rm R},\kappa)=(-1,-1,1)\omega_{\rm rec}$.}
\label{fig:phase_diag} 
\end{figure}

As the frequency and the strength of the pump lasers are varied, in the mean-field regime the system displays three fundamentally different phases with distinct characteristics in density and spatial spin texture as shown in~\fref{fig:phase_diag}. The first phase is the \textit{density-wave--spin-wave} (DW-SW) state, where the density has a crystalline structure and the pseudospin exhibits a spatial spin-wave texture along the cavity axis; see \fref{fig:spin}(a). The second phase is the \textit{plane-wave--spin-spiral} (PW-SS) state, where the density is homogeneous while the pseudospin exhibits a spin-spiral texture as illustrated in \fref{fig:spin}(b). The third phase is the \textit{density-wave--spin-spiral} (DW-SS) state, where a crystalline-ordered density is accompanied with a spin-spiral pseudospin texture; see \fref{fig:spin}(c). The latter two states with the spin-spiral texture have a Skyrmionic nature with a nontrivial topology~\cite{nagaosa2013topological,fert2013skyrmions}, and are intimately related to the emergence of cavity-induced spin-orbit coupling. The topologically trivial DW-SW state, on the other hand, breaks the continuous screw-like symmetry of the system, resulting in the appearance of a gapless Goldstone mode in addition to the phonon sound mode, a characteristic of a supersolid with two broken continuous symmetries~\cite{leonard2017supersolid,Li2017stripe,Mivehvar2018}. The topological phase transitions between the DW-SW and the spin-spiral states, which is a direct consequence of the emergence of cavity-induced spin-orbit coupling, exhibit first-order characteristics, while the quantum phase transition between the two spin-spiral states are second order. Remarkably, all the quantum phase transitions can be monitored \textit{in situ} through the cavity output, as can be seen from the inset of~\fref{fig:phase_diag}.

The paper is organized as follows. We introduce the model in Sec.~\ref{sec:model} and then derive the effective Hamiltonian and the Heisenberg equations of motion. We then find the steady-state solutions of the equations of motion in the mean-field limit in Sec.~\ref{sec:phase_diagram}. In this section we discuss the atomic phase diagram (Sec.~\ref{subsec:densmod}), the cavity-field phase diagram (Sec.~\ref{subsec:fields}), and the effect of cavity-induced emergent spin-orbit coupling (Sec.~\ref{subsec:SOC}). Section~\ref{sec:coll_ex} is devoted to the elementary excitations and the (broken) symmetries of the system. We present the concluding remarks in Sec.~\ref{sec:conclusion_and_outlook}. Appendices~\ref{app:A} and \ref{app:B} show the details of the adiabatic elimination of the atomic excited state and linearization of the Heisenberg equations of motion, respectively.

\section{Model}
\label{sec:model}
Consider a $\Lambda$-type spin-1 BEC tightly confined into quasi-one dimension along one leg of a ring cavity in $z$ direction as depicted in~\fref{fig:setup}. The internal atomic states of interest consist of two pseudospin ground states, designated by $\ket{\downarrow}$ and $\ket{\uparrow}$, and an electronic excited state $\ket{e}$ with energies $\{\hbar\omega_\downarrow,\hbar\omega_\uparrow,\hbar\omega_e\}$. The transition $\ket{\downarrow}\leftrightarrow\ket{e}$ ($\ket{\uparrow}\leftrightarrow\ket{e}$) couples to a pair of degenerate, counterpropagating electromagnetic modes $\hat{a}_\pm e^{\pm ik_az}$ ($\hat{b}_\pm e^{\pm ik_bz})$ of the ring cavity as shown in~\fref{fig:setup}. The operator $\hat{a}_{+/-}$ ($\hat{b}_{+/-}$) annihilates a forward/backward moving photon in the first (second) pair of ring-cavity modes with a wave vector $k_a=\omega_a/c=2\pi/\lambda_a$ ($k_b=\omega_b/c=2\pi/\lambda_b$). For our desired system, the condition $|\omega_a-\omega_b|/\omega_{a(b)}\ll1$ holds in general. Therefore, the assumptions $k\coloneqq k_a\approx k_b$ and $\lambda\coloneqq\lambda_a~\approx\lambda_b$ are legitimate and will be used throughout this work. Each pair of modes (i.e., $\hat{a}_\pm$ or $\hat{b}_\pm$) has the same polarization, which is orthogonal to the polarization of the other pair; e.g., $\{\epsilon_+,\epsilon_-\}$. The mode $\hat{a}_+$ ($\hat{b}_-$) is driven by an external pump laser through the cavity mirror with a frequency $\tilde\omega_a$ ($\tilde\omega_b$) and an amplitude $\eta_+$ ($\eta_-$).

The single-particle Hamiltonian density of the system in the dipole and rotating-wave approximation reads
\begin{align} \label{eq:total-single-atom-H-density}
\mathcal{H}=\mathcal{H}_{0,\rm at}+H_{0,\rm cav}+\mathcal{H}_{\rm ac}+H_{\rm pump},
\end{align}
with
\begin{align} 
\mathcal{H}_{0,\rm at}&=\frac{\hat{p}^2}{2m}
+\hbar\sum_{\tau=\downarrow,\uparrow,e}\omega_\tau \hat{\sigma}_{\tau \tau},\nonumber\\
H_{0,\rm cav}&=\hbar\sum_{j=+,-} 
\left(\omega_a\hat{a}^\dag_j \hat{a}_j+\omega_b\hat{b}^\dag_j \hat{b}_j\right),\nonumber\\
\mathcal{H}_{\rm ac}&=\hbar 
\Big\{\Big[\mathcal{G}_\downarrow\left(e^{ikz}\hat{a}_++e^{-ikz}\hat{a}_-\right)\hat{\sigma}_{e\downarrow}\nonumber\\
&~~\qquad+\mathcal{G}_\uparrow\left(e^{ikz}\hat{b}_++e^{-ikz}\hat{b}_-\right)\hat{\sigma}_{e\uparrow}\Big] +\text{H.c.} \Big\},\nonumber\\
H_{\rm pump}&=i\hbar\left(\eta_+ \hat{a}_+^\dag e^{-i\tilde\omega_at} + \eta_- \hat{b}_-^\dag e^{-i\tilde\omega_bt} - \text{H.c.}\right).
\end{align}
Here $m$ is the atomic mass, $\hat{p}=i\hbar\partial_z$ is the atomic momentum operator and $\hat{\sigma}_{\tau \tau'}=\ket{\tau}\bra{\tau'}$ are the atomic transition operators. The atom-photon coupling for the transition $\ket{\downarrow}\leftrightarrow \ket{e}$ ($\ket{\uparrow}\leftrightarrow \ket{e}$) is denoted as $\mathcal{G}_\downarrow$ ($\mathcal{G}_\uparrow$) and H.c.\ stands for the Hermitian conjugate. $\mathcal{H}_{0,\rm at}$ and $H_{0,\rm cav}$ are the bare atomic Hamiltonian density and the cavity-field Hamiltonian, respectively. $\mathcal{H}_{\rm ac}$ represents the coupling between the atom and the cavity fields, and $H_{\rm pump}$ accounts for the pumping of the cavity fields.

The corresponding many-body Hamiltonian is obtained as 
\begin{align}
H&=\int \hat\Psi^\dag(z)(\mathcal{H}_{0,\rm at}+\mathcal{H}_{\rm ac})\hat\Psi(z) dz\\
&+H_{0,\rm cav}+H_{\rm pump}+H_{\rm int}\nonumber, 
\end{align}
where $\hat\Psi=(\hat\psi_e,\hat\psi_\uparrow,\hat\psi_\downarrow)^\top$ is the three-component atomic field operator which satisfies the bosonic commutation relation $[\hat\psi_\tau(z),\hat\psi_{\tau'}^\dag(z')]=\delta(z-z')\delta_{\tau,\tau'}$. The Hamiltonian $H_{\rm int}$ accounts for two-body contact interactions between the atoms, and ensures the thermalization and relaxation of the BEC. However, we assume that the two-body contact interactions are negligibly small compared to cavity-mediated long-range interactions, which is a good approximation for typical cavity-QED experiments~\cite{leonard2017supersolid,schuster2018pinning}. Therefore, we will not explicitly include two-body contact interactions in our model.  
 
The dynamics of the system is governed by the Heisenberg equations of motion of the atomic field operators $i\hbar\partial_t\hat{\psi}_\tau=[\hat{\psi}_\tau,H]$, and the photonic field operators $i\hbar\partial_t\hat{a}_\pm=[\hat{a}_\pm,H]-i\hbar\kappa\hat{a}_\pm$ and $i\hbar\partial_t\hat{b}_\pm=[\hat{b}_\pm,H]-i\hbar\kappa\hat{b}_\pm$. The decay (i.e., leakage) of cavity photons is included phenomenologically by adding the terms proportional to $\kappa$ in the latter equations for the photonic-field operators. If the relative atomic detunings with respect to the pump lasers $\Delta_{\downarrow(\uparrow)}\coloneqq\tilde\omega_{a(b)}-[\omega_e-\omega_{\downarrow(\uparrow)}]$ are large compared to the two-photon detuning $\delta\coloneqq\Delta_\uparrow-\Delta_\downarrow$ and the atom-photon couplings $\{\mathcal{G}_\downarrow,\mathcal{G}_\uparrow\}$, the atomic excited state reaches a steady-state on a short time scale and its dynamics can be adiabatically eliminated. This results in  a set of six coupled effective Heisenberg equations for the atomic pseudospin and photonic field operators
\begin{subequations} \label{eq:app-coupled_eff_Heisenberg_eqs}
\begin{align} 
i\hbar\frac{\partial}{\partial t}
\begin{pmatrix}
\hat\psi_\downarrow \\ 
\hat\psi_\uparrow \\
\end{pmatrix}&=
{\mathcal H}_{\rm at}
\begin{pmatrix}
\hat\psi_\downarrow \\ 
\hat\psi_\uparrow \\
\end{pmatrix},
\label{eq:app-coupled_eff_Heisenberg_eqs-atom}
\\
i\hbar\frac{\partial}{\partial t}
\begin{pmatrix}
\hat{a}_+ \\
\hat{a}_- \\
\hat{b}_+ \\
\hat{b}_-
\end{pmatrix}&=\mathcal{H}_{\rm cav}
\begin{pmatrix}
\hat{a}_+ \\
\hat{a}_- \\
\hat{b}_+ \\
\hat{b}_-
\end{pmatrix}
+i\hbar
\begin{pmatrix}
\eta_+ \\
0 \\
0 \\
\eta_-
\end{pmatrix}.
\label{eq:app-coupled_eff_Heisenberg_eqs-cavity}
\end{align}
\end{subequations}
The details are presented in Appendix~\ref{app:A}.

Now ${\mathcal H}_{\rm at}$ and $\mathcal{H}_{\rm cav}$ are the ``effective'' atomic and cavity-field Hamiltonian densities, respectively, which contain the couplings between all atomic and photonic degrees of freedom. The effective atomic Hamiltonian density has the matrix form 
\begin{align} \label{eq:eff-atomic-H-density}
{\mathcal H}_{\rm at}=
\begin{pmatrix}
\frac{\hat{p}^2}{2m}+\hbar\hat{U}_\downarrow(z)-\frac{\hbar\delta}{2} & 
\hbar\hat{\Omega}_{\rm R}(z) \\ 
\hbar\hat{\Omega}_{\rm R}^\dag(z) &
\frac{\hat{p}^2}{2m}+\hbar\hat{U}_\uparrow(z)+\frac{\hbar\delta}{2} \\
\end{pmatrix},
\end{align}
with the potential operators 
\begin{align}\label{eq:pot-oper}
\hat{U}_\downarrow(z)&=U_{0\downarrow} \big(\hat{a}_+^\dag\hat{a}_++\hat{a}_-^\dag\hat{a}_-
+e^{-2ikz}\hat{a}_+^\dag\hat{a}_-+e^{2ikz}\hat{a}_-^\dag\hat{a}_+\big),\nonumber\\
\hat{U}_\uparrow(z)&=U_{0\uparrow} \big(\hat{b}_+^\dag\hat{b}_++\hat{b}_-^\dag\hat{b}_-
+e^{-2ikz}\hat{b}_+^\dag\hat{b}_-+e^{2ikz}\hat{b}_-^\dag\hat{b}_+\big),
\end{align}
and the two-photon Raman coupling operator
\begin{align} \label{eq:Raman-oper}
\hat{\Omega}_{\rm R}(z)=\Omega_{\rm 0R}\big(\hat{a}_+^\dag\hat{b}_++\hat{a}_-^\dag\hat{b}_-
+e^{-2ikz}\hat{a}_+^\dag\hat{b}_-+e^{2ikz}\hat{a}_-^\dag\hat{b}_+\big).
\end{align}

Here we have introduced the maximum depth of the optical potential per photon $U_{0\tau}\coloneqq2|\mathcal{G}_{\tau}|^2/(\Delta_\downarrow+\Delta_\uparrow)$ and the maximum two-photon Raman transition frequency $\Omega_{\rm 0R}\coloneqq2\mathcal{G}_\downarrow^*\mathcal{G}_\uparrow/(\Delta_\downarrow+\Delta_\uparrow)$. The former potential depth results from two-photon scatterings between cavity modes with the same polarization without changing the atomic internal state, whereas the latter frequency $\Omega_{\rm 0 R}$ is due to two-photon scatterings between cavity modes with orthogonal polarizations accompanied by an atomic pseudospin flip $\ket{\downarrow} \leftrightarrow \ket{\uparrow}$.

\begin{figure*}
\centering
\includegraphics[width=0.8\textwidth]{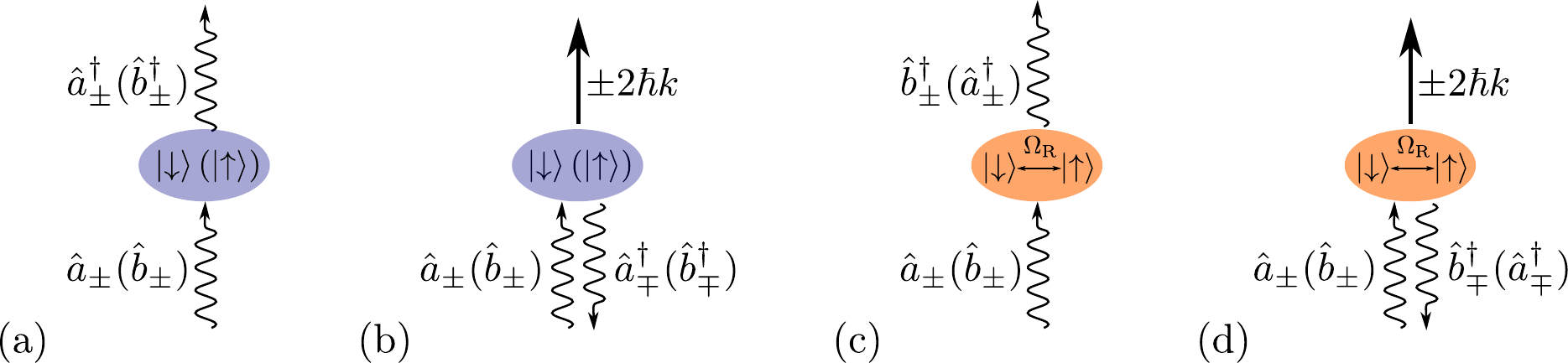}
\caption{Schematic visualization of the two-photon Raman processes. 
The coherent scattering of a photon from a mode into itself (a) and into the 
corresponding degenerate counterpropagating mode with the same polarization (b)
without changing the atomic internal state result in position-independent and position-dependent potential terms in $\hat{U}_\tau(z)$. The coherent scattering of a photon from a mode into another mode with an orthogonal polarization propagating in the same (c) and opposite (d) direction, accompanied with the atomic pseudospin flip $\ket{\downarrow} \leftrightarrow \ket{\uparrow}$, gives rise to position-independent and position-dependent Raman coupling terms in $\hat{\Omega}_{\rm R}(z)$, respectively.}
\label{fig:processes}
\end{figure*}

The coherent scattering of photons from a mode back into itself results in the position-independent energy shifts (\ie, the terms $\hat{a}_\pm^\dag\hat{a}_\pm$ and $\hat{b}_\pm^\dag\hat{b}_\pm$) in the potential operators $\hat{U}_\tau(z)$, while the scattering of photons from a mode (say, $\hat{a}_+$) into its degenerate counterpropagating mode (i.e., $\hat{a}_-$) results in the position-dependent terms proportional to $e^{\pm2ikz}$ in the potential operators. The former photon scatterings do not transfer any momentum to the atom, while the latter scatterings lead to $\pm2\hbar k$ momentum kicks to the atom. These particular processes are schematically shown in~\fref{fig:processes}(a) and~\fref{fig:processes}(b), respectively.

The photon scattering between modes with orthogonal polarizations and the same propagation direction (that is, between $\hat{a}_\pm$ and $\hat{b}_\pm$) gives rise to the position-independent Raman coupling terms (i.e., the terms $\hat{a}_+^\dag\hat{b}_+$ and $\hat{a}_-^\dag\hat{b}_-$) in $\hat{\Omega}_{\rm R}(z)$, while scattering between modes with orthogonal polarizations and opposite propagation directions (that is, between $\hat{a}_\pm$ and $\hat{b}_\mp$) results in the position-dependent Raman coupling terms proportional to $e^{\pm2ikz}$ in the two-photon Raman coupling operator. These two-photon Raman processes are illustrated in \fref{fig:processes}(c) and \fref{fig:processes}(d). While the atomic pseudospin is flipped in both Raman processes, there is no net momentum transfer to the atom in the former processes, whereas the latter processes impart a $\pm2\hbar k$ momentum to the atom. These latter Raman transition processes with $\pm2\hbar k$ momentum kicks can induce a synthetic spin-orbit coupling for the atom.

The effective cavity-field Hamiltonian ``density'' in matrix form is given by
\begin{align} \label{eq:app-dynamic_matrix}
\mathcal{H}_{\rm{cav}}=
\renewcommand\arraystretch{1.5}
\hbar
\begin{pmatrix}
-\tilde{\Delta}_a & U_{0\downarrow}  \hat{\mathcal N}_\downarrow & 
\Omega_{\rm0R}\hat{S}_- & \Omega_{\rm0R}\hat{\mathcal S}_-^{(-)}
\\
U_{0\downarrow}\hat{\mathcal N}_\downarrow^\dag & -\tilde{\Delta}_a &
\Omega_{\rm0R}\hat{\mathcal S}_-^{(+)} & \Omega_{\rm0R}\hat{S}_-
\\
\Omega_{\rm0R}^*\hat{S}_+ &  \Omega_{\rm0R}^*\hat{\mathcal S}_+^{(-)} & 
-\tilde{\Delta}_b & U_{0\uparrow}\hat{\mathcal N}_\uparrow 
\\
\Omega_{\rm0R}^*\hat{\mathcal S}_+^{(+)} & \Omega_{\rm0R}^*\hat{S}_+ & 
U_{0\uparrow}\hat{\mathcal N}^\dag_\uparrow & -\tilde{\Delta}_b 
\end{pmatrix},
\end{align} 
where we have introduced the effective cavity detunings $\tilde{\Delta}_{a(b)}\coloneqq(\Delta_{a(b)}+i\kappa)-U_{0\downarrow(\uparrow)}\hat{N}_{\downarrow(\uparrow)}$ with the relative cavity detunings with respect to the pump frequencies $\Delta_{a(b)}\coloneqq\tilde\omega_{a(b)}-\omega_{a(b)}$ and the particle number operators
\begin{equation}\label{eq:part_numb_operator}
\hat{N}_{\tau}=\int  \hat{\psi}^\dag_{\tau}(z)\hat{\psi}_{\tau}(z) dz.
\end{equation}
The off-diagonal coupling operators are given by
\begin{subequations}\label{eq:app-dynamic_matrix_elements-cavity}
\begin{align}
\hat{\mathcal N}_{\tau}&=\int e^{-2ik z}\hat{\psi}^\dag_{\tau}(z)\hat{\psi}_{\tau}(z)  dz,\label{eq:order_par}\\
\hat{S}_-&=\hat{S}_+^\dag=\int  \hat{\psi}_\downarrow^\dag(z)\hat{\psi}_\uparrow(z) dz,\label{eq:Sminus}\\
\hat{\mathcal S}_{-}^{(\pm)}&=(\hat{\mathcal S}_{+}^{(\mp)})^\dag=\int e^{\pm 2ikz} \hat{\psi}_\downarrow^\dag(z)\hat{\psi}_\uparrow(z)dz, \label{eq:curlSminus}
\end{align}
\end{subequations}
where $\hat{S}_+$ ($\hat{S}_-$) is the collective atomic spin raising (lowering) operator. The operators $\hat{\mathcal N}_{\tau}$ and $\hat{\mathcal S}_{\pm}^{(\pm)}$ are the density- (for pseudospin $\tau$) and spin-wave operators, respectively.

The matrix elements of the effective cavity-field Hamiltonian density $\mathcal{H}_{\rm cav}$ give the strengths of the two-photon processes depicted in~\fref{fig:processes}. The diagonal terms proportional to $\hat{N}_\tau$ correspond to the processes illustrated in~\fref{fig:processes}(a) and result in the dispersive shifts $U_{0\tau}\hat{N}_\tau$ of the cavity frequencies. Terms proportional to $\hat{\mathcal{N}}_\tau$ (and their Hermitian conjugates) correspond to the processes shown in~\fref{fig:processes}(b) and provide the couplings between degenerate counterpropagating modes with the same polarization. Finally, the matrix elements proportional to $\hat{S}_\pm$ and $\hat{\mathcal{S}}_\pm^{(\pm)}$ correspond to the pseudospin flipping processes depicted in~\fref{fig:processes}(c) and~\fref{fig:processes}(d), respectively, and yield the nontrivial couplings between modes with orthogonal polarizations.

The system possesses a screw-like continuous symmetry. It is manifested in the invariance of the total effective Hamiltonian corresponding to Eqs.~\eqref{eq:app-coupled_eff_Heisenberg_eqs} under a simultaneous spatial translation $z\rightarrow z+\Delta z$, phase rotations of the unpumped photonic field operators $\hat{a}_-\rightarrow\hat{a}_-e^{2ik\Delta z}$ and $\hat{b}_+\rightarrow\hat{b}_+e^{-2ik\Delta z}$, and phase rotations of the atomic field operators $\hat{\psi}_\downarrow\rightarrow\hat{\psi}_\downarrow e^{-ik\Delta z}$ and $\hat{\psi}_\uparrow\rightarrow\hat{\psi}_\uparrow e^{+ik\Delta z}$. Note that the phases of the pumped cavity modes $\hat{a}_+$ and $\hat{b}_-$ are fixed by the cavity pumps $\eta_\pm$. The phase rotation of $\hat{a}_-$ and $\hat{b}_+$ results in a shift of the potential minima in~\eref{eq:pot-oper}, defining the position of the atomic-density maxima. On the other hand, the phase shift of the atomic field operators leads to rotations of the spin operators: $\hat{S}_+\rightarrow \hat{S}_+ e^{-2ik\Delta z}$ and $\hat{\mathcal{S}}^{(\pm)}_+\rightarrow \hat{\mathcal{S}}^{(\pm)}_+ e^{-2ik\Delta z}$. Hence, a translation in space is tied to a corresponding rotation of the atomic spin, where the rotation angle directly depends on the length of the spatial translation $\Delta z$. This leads to a screw-like continuous symmetry.

The system is highly nonlinear. The effective atomic Hamiltonian density depends on the cavity fields through the potential $\hat{U}_\tau(z)$ and the Raman operators $\hat{\Omega}_{\rm R}(z)$, while the effective cavity-field Hamiltonian density depends on the atomic fields via the atomic number $\hat{N}_\tau$, the density-wave $\hat{\mathcal{N}}_\tau$, the collective atomic spin $\hat{S}_\pm$, and the spin-wave $\hat{\mathcal S}_{\pm}^{(\pm)}$ operators. It is this nonlinear dynamics and the nontrivial interplay between various degrees of freedom which give rise to intriguing phenomena in our system, as it will be discussed in the subsequent sections. These nonlinear dynamics and the nontrivial interplay between the cavity modes survive even in the strong pumping limit $\eta_\pm\gg\omega_{\rm rec}$, with $\omega_{\rm rec}\coloneqq\hbar k^2/2m$ being the recoil frequency. Although in the strong pumping limit the pumped cavity fields $\{\hat{a}_+,\hat{b}_-\}$ behave as classical fields, the unpumped modes $\{\hat{a}_-,\hat{b}_+\}$ still retain their quantum nature and behave as dynamical fields.

\begin{figure*}[t!]
\centering
(a) \includegraphics[width=0.29\textwidth]{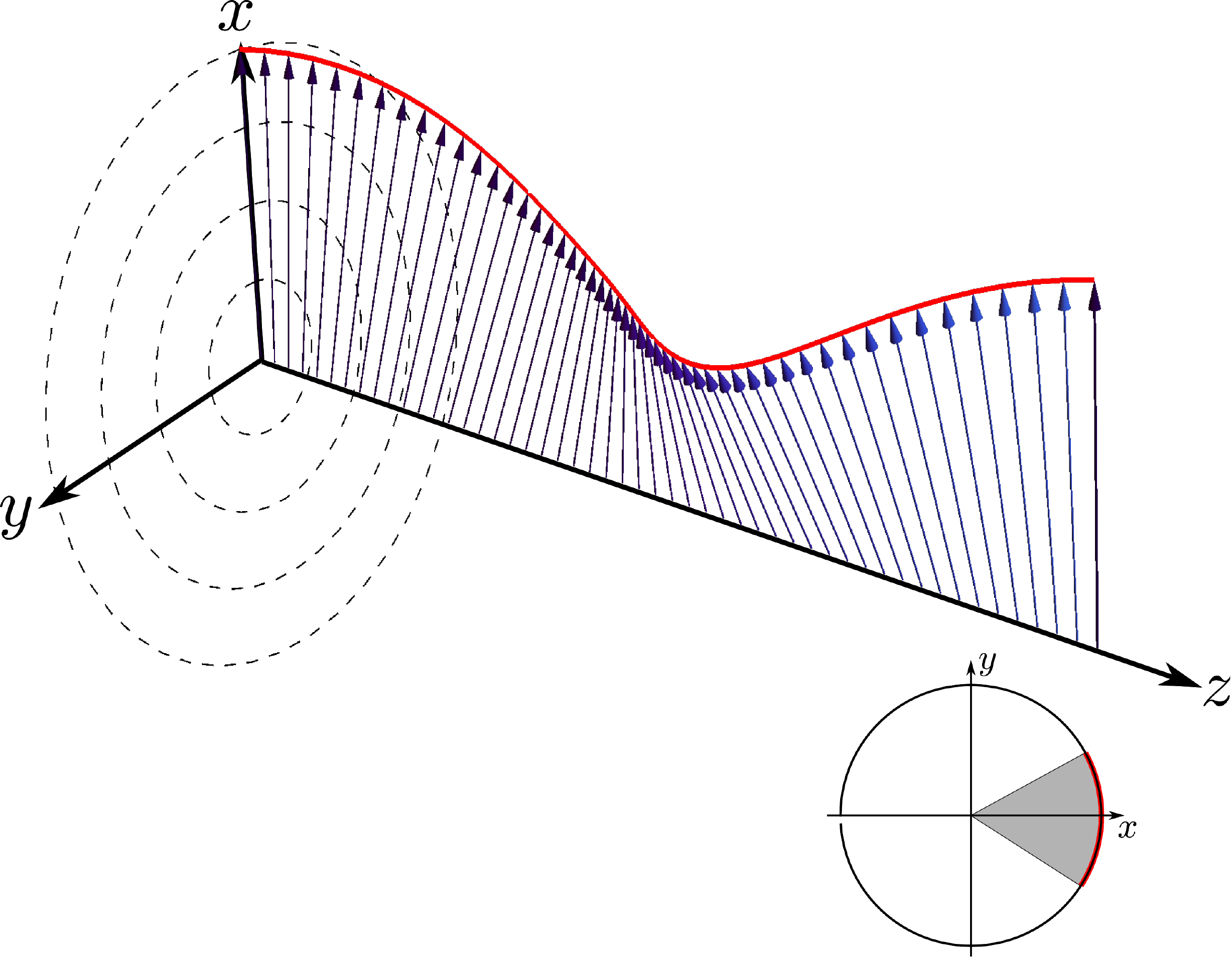}
\hspace{0.7cm}
(b) \includegraphics[width=0.23\textwidth]{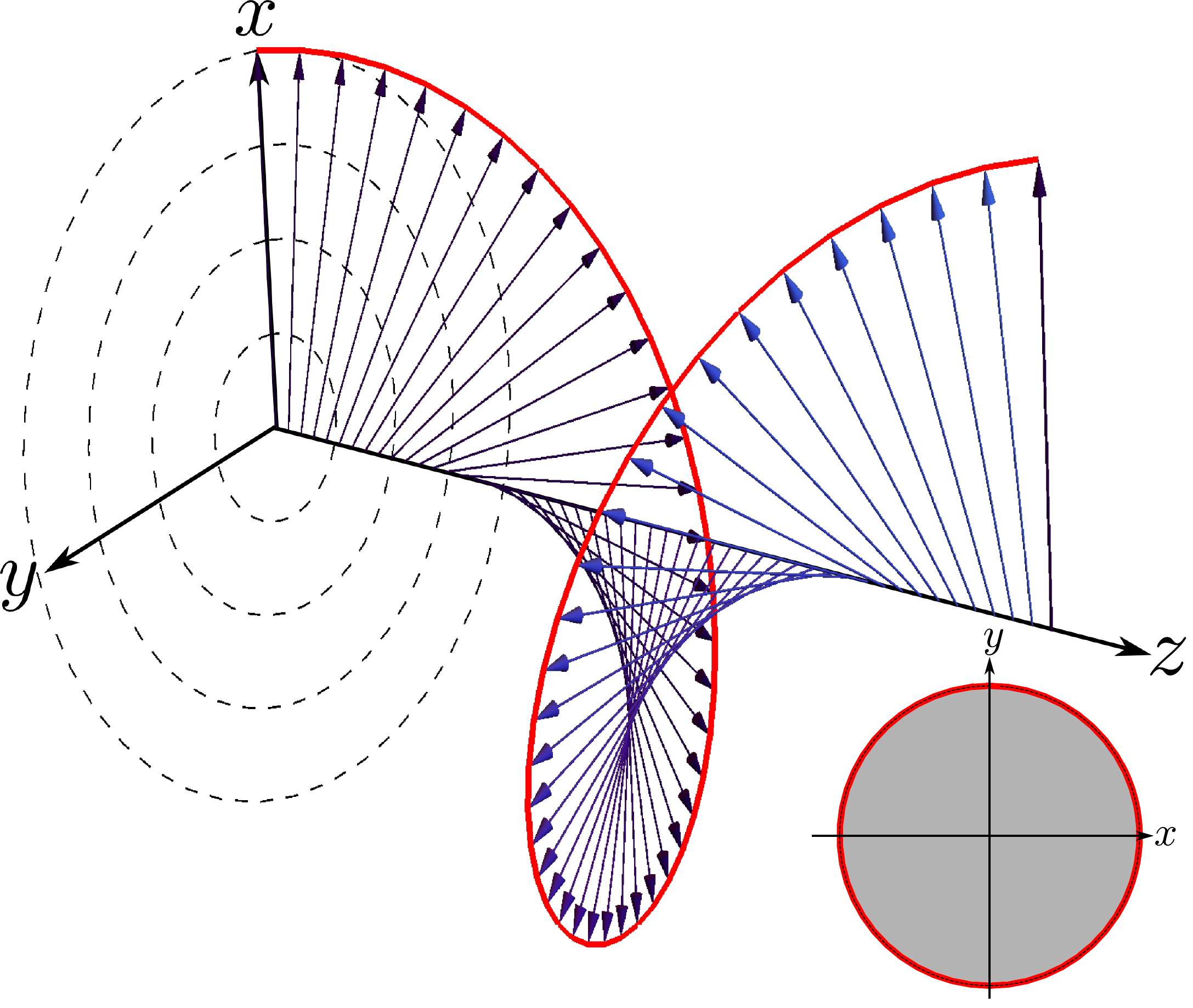}
\hspace{0.7cm}
(c) \includegraphics[width=0.23\textwidth]{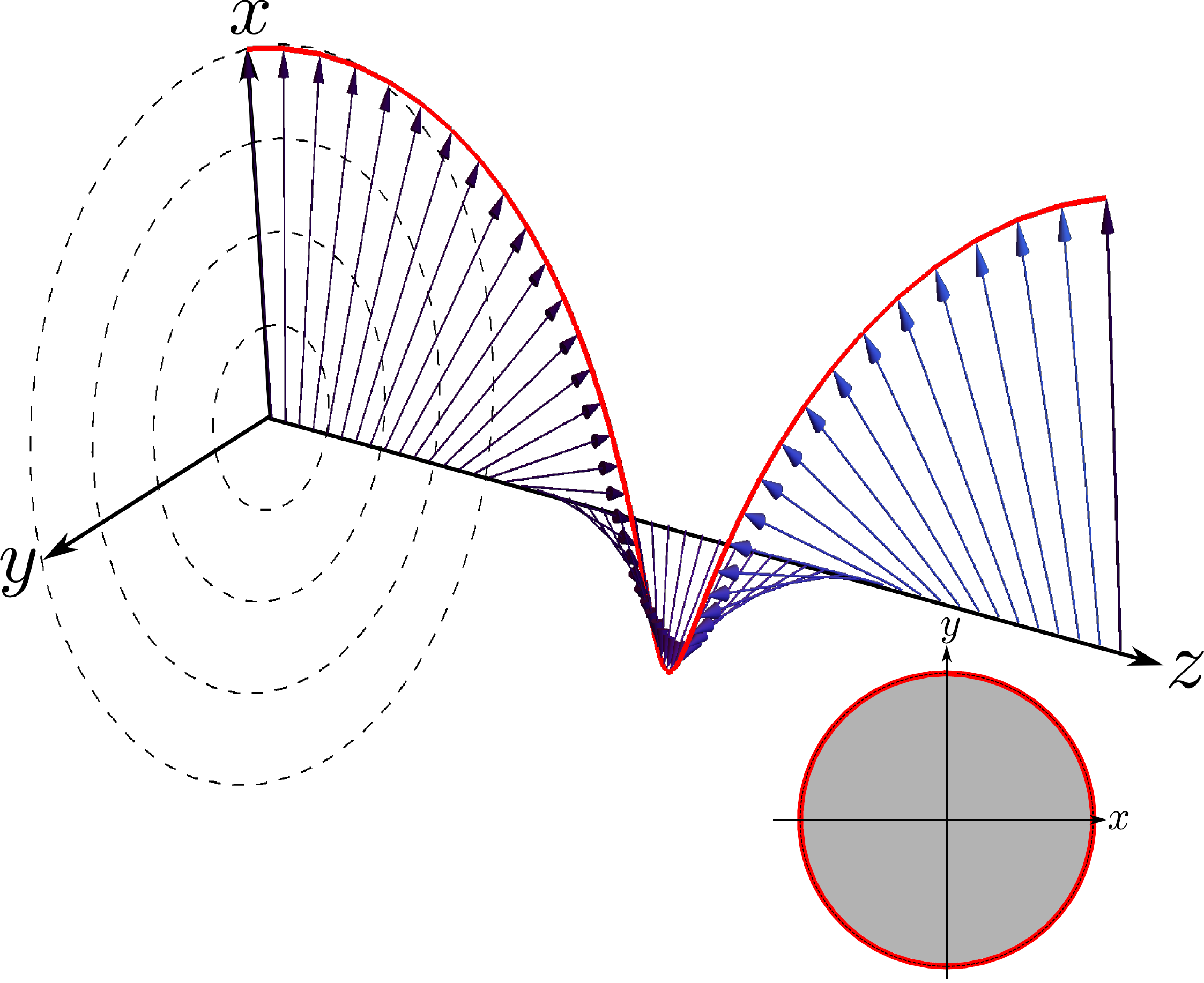}
\caption{Spin textures along the cavity axis $z$ in one unit cell of length $\lambda/2$ in different phases. 
The local atomic pseudospin vector $\mathbf{s}(z)$ is shown 
in the DW-SW phase for the parameters $(\Delta,\sqrt{N}\eta)=(-20,20)\omega_\mathrm{rec}$ (a),
the PW-SS phase for $(\Delta,\sqrt{N}\eta)=(-20,30)\omega_\mathrm{rec}$ (b),
and the DW-SS phase for $(\Delta,\sqrt{N}\eta)=(-20,50)\omega_\mathrm{rec}$ (c).
The small circles in the lower right corners display the projection of the spin textures in the $x$-$y$ plane,
where the grey regions indicate the angles swept by the pseudospin vector over a $\lambda/2$ distance.
The other parameters are the same as \fref{fig:phase_diag}.}
\label{fig:spin}
\end{figure*}

\section{Mean-field results}
\label{sec:phase_diagram}

In the following, we restrict our analysis to red-detuned pump lasers with respect to both bare atomic and cavity frequencies, i.e., $\{\Delta_{\downarrow,\uparrow},\Delta_{a,b}\}<0$. The atoms are therefore attracted to the intensity maxima of the light fields, while experiencing cavity cooling. In order to reduce the number of free parameters and capture the fundamental physics, we further focus on the special case of a completely symmetric configuration: $\Delta\coloneqq\Delta_a=\Delta_b$, $\eta\coloneqq\eta_+=\eta_-$, $\delta=0$, and $\mathcal{G}_\downarrow=\mathcal{G}_\uparrow$ which results in $U_0\coloneqq U_{0\downarrow}= U_{0\uparrow}=\Omega_{\rm0R}$. Despite these simplifying assumptions, the system is still very complex and gives rise to intriguing phenomena.

We find the stationary states of the system by self-consistently solving Eqs.~\eqref{eq:app-coupled_eff_Heisenberg_eqs}-\eqref{eq:app-dynamic_matrix_elements-cavity} in the mean-field regime in the parameter space $\{\eta,\Delta\}$. This amounts to omitting quantum fluctuations and replacing the atomic and cavity field operators by their corresponding quantum averages: $\hat{\psi}_\tau\rightarrow\psi_\tau\coloneqq \langle\hat{\psi}_\tau\rangle$, $\hat{a}_j\to\alpha_j\coloneqq\langle\hat{a}_j\rangle$, and $\hat{b}_j\to\beta_j\coloneqq\langle\hat{b}_j\rangle$. The parameters $\eta$ and $\Delta$ are related, respectively, to the intensity and the frequency of the external pump lasers and can be readily tuned in experiment.

The effective atomic Hamiltonian density~\eqref{eq:eff-atomic-H-density} is $\lambda/2$ periodic. Nonetheless, solving the equations for different numbers of unit cells (of length $\lambda/2$) reveals that the atomic condensate wave functions $\psi_\tau(z)$ are $\lambda$ periodic in parameter regimes possessing cavity-induced spin-orbit coupling. Therefore, we always solve the mean-field equations corresponding to Eqs.~\eqref{eq:app-coupled_eff_Heisenberg_eqs}-\eqref{eq:app-dynamic_matrix_elements-cavity} in two unit cells of total length $\lambda$ with periodic boundary conditions.  The relation between cavity-induced spin-orbit coupling and the doubling of the periodicity of the condensate wave functions will be discussed in more details in Sec.~\ref{subsec:SOC}.

\subsection{Atomic phase diagram}
\label{subsec:densmod}

The mean-field density-wave order parameters $\mathcal{N}_\tau=\langle\hat{\mathcal N}_\tau\rangle$, cf.\ Eq.~\eqref{eq:order_par}, can be used to characterize the density structure of each BEC component. They quantify the magnitude of the density modulations of each BEC component, where a zero density-wave order parameter corresponds to a homogenous density distribution. Because of the symmetric choice of the parameters as described above, we always find that the absolute values of the two density-wave order parameters are equal to one another, $|\mathcal{N}_\downarrow|=|\mathcal{N}_\uparrow|$.

On the other hand, the mean-field local pseudospin vector $\mathbf{s}(z)=(s_x(z),s_y(z),s_z(z))=\bra{\psi_\mathrm{eff}(z)}\pmb\sigma\ket{\psi_\mathrm{eff}(z)}$, where $\ket{\psi_\mathrm{eff}(z)}:=(\psi_\uparrow(z),\psi_\downarrow(z))^\top$ and $\pmb\sigma=(\sigma_x,\sigma_y,\sigma_z)$ is the vector of the Pauli matrices, can be used to illustrate the spatial spin texture of the steady-states. The $z$ component of the local pseudospin $s_z(z)=[|\psi_\uparrow(z)|^2-|\psi_\downarrow(z)|^2]/2$ is zero everywhere in all parameter regimes due to the symmetric choice of the parameters. Therefore, the local pseudospin vector always lies in the $x$-$y$ plane. We find that the transverse local pseudospin vector varies in space in all parameter regimes. In some regimes $\mathbf{s}(z)$ exhibits a $\lambda/2$-periodic ``spin wave'' of ferromagnetic-magnon nature, meaning that the spin angle $\phi(z):=\arctan(s_y(z)/s_x(z))$ only sweeps a small sector within the interval $[0,\pi/2]$ over a $\lambda/2$ distance. While for other parameter regimes, $\phi(z)$ sweeps a full $2\pi$ angle in the $x$-$y$ plane over a $\lambda/2$ distance, leading to a $\lambda/2$-periodic ``spin spiral'' of topological Skyrmionic nature~\cite{nagaosa2013topological,fert2013skyrmions}. As a result, the mean-field collective atomic pseudospins $S_\pm=\langle\hat{S}_\pm\rangle=\int [s_x(z)\pm is_y(z)] dz$ and the spin-wave order parameters $\mathcal{S}_\pm^{(\pm)}=\langle\hat{\mathcal S}_\pm^{(\pm)}\rangle=\int e^{\pm2ikz}[s_x(z)\pm is_y(z)] dz$, cf.\ Eqs.~\eqref{eq:Sminus} and \eqref{eq:curlSminus}, exhibit different behaviors in the spin-wave and the spin-spiral states.

The spin-wave and spin-spiral states can be quantitatively distinguished by their distinct topological structures via an appropriate topological invariant. The relevant topological invariant to characterize the spin texture of the system is the winding number~\cite{asboth2016short,Braun2012}
\begin{align}\label{eq:wind_numb}
\mathcal{W}&\coloneqq\frac{1}{2\pi}\int_0^{\lambda/2} [\partial_z \phi(z)] dz=\frac{\phi(\lambda/2)-\phi(0)}{2\pi},
\end{align}
where $\phi(z)$ defines the direction of the local pseudospin vector $\mathbf{s}(z)$ in the $x$-$y$ plane. Note that the angle $\phi(z)$ is tied to the relative phase between the two condensate wave functions $\psi_\downarrow(z)$ and $\psi_\uparrow(z)$. The winding number essentially counts the number of full rotations of the local pseudospin vector $\mathbf{s}(z)$ around the origin in one unit cell. Zero winding number corresponds to the topologically trivial spin-wave state, while a nonzero winding number (i.e., $\mathcal{W}=+1$)  indicates the topological spin-spiral state.

\begin{figure*}[t!]
\center{
(a) \includegraphics[width=0.45\textwidth]{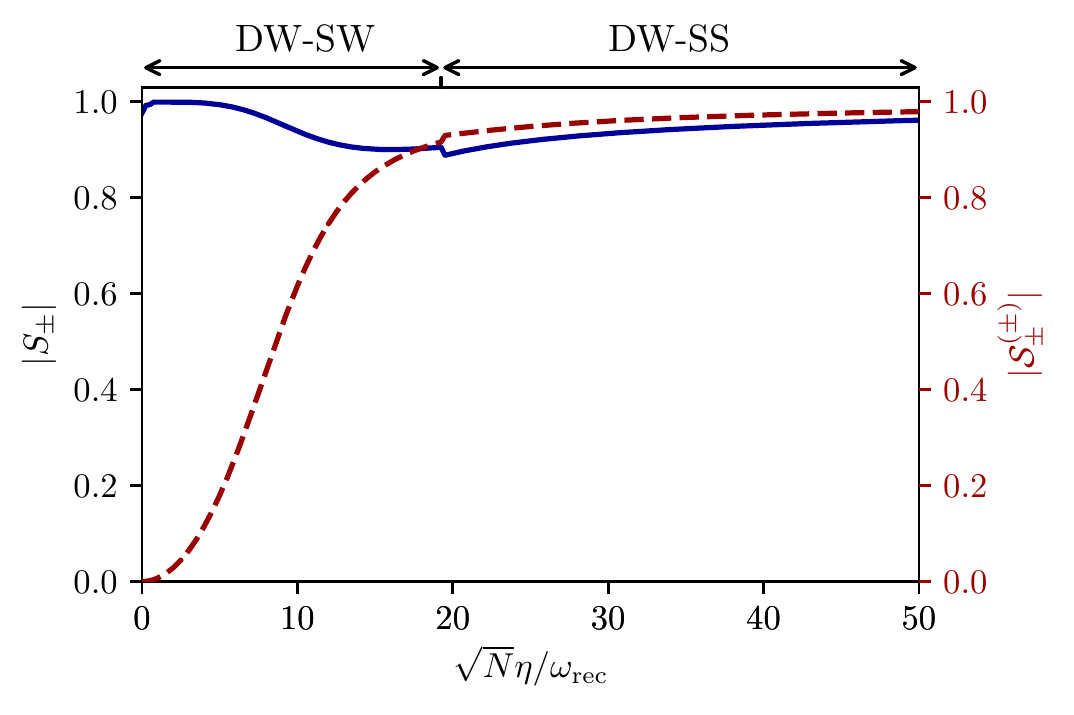}
(b) \includegraphics[width=0.45\textwidth]{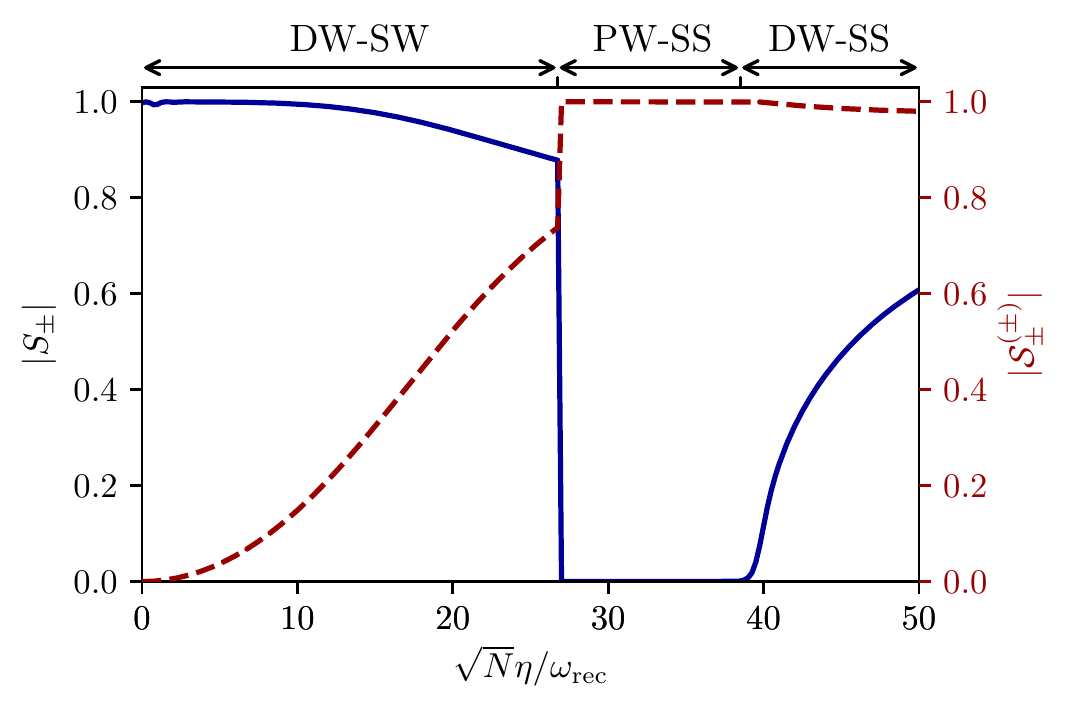}}
\caption{
The absolute values of the collective atomic spins $|S_\pm|$ (blue solid curve) and the spin-wave order parameters $|\mathcal{S}_\pm^{(\pm)}|$ (red dashed curve) as a function of the rescaled pump strength $\sqrt{N}\eta/\omega_{\rm rec}$ for $\Delta=-10\omega_{\rm rec}$ (a) and $\Delta=-20\omega_{\rm rec}$ (b). The collective atomic spins $S_\pm$ are the dominant parameters in the DW-SW phase, while the spin-wave order parameters $\mathcal{S}_\pm^{(\pm)}$ are dominant in the PW-SS and DW-SS phases. Both parameters $\{S_\pm,\mathcal{S}_\pm^{(\pm)}\}$ exhibit first-order (second-order) characteristics across the phase transitions between the DW-SW and the DW-SS/PW-SS (between the PW-SS and the DW-SS). The other parameters are the same as \fref{fig:phase_diag}.}
\label{fig:spin-comp-vs-eta}
\end{figure*}

The atomic phase diagram of the system in the rescaled parameter space $\{\sqrt{N}\eta/\omega_{\rm rec},\Delta/\omega_{\rm rec}\}$ is shown in~\fref{fig:phase_diag} and displays three fundamentally different phases. In the first phase,  which corresponds to the region below the solid red curve in the phase diagram, the density order parameters are nonzero, $|\mathcal{N}_\tau|\neq0$, while the winding number is zero, $\mathcal{W}=0$. This implies that the atomic density distribution $n_\tau(z)=|\psi_\tau(z)|^2$ has a ($\lambda/2$-periodic) crystalline order and the local pseudospin $\mathbf{s}(z)$ exhibits a (also $\lambda/2$-periodic) spin-wave texture. Hence, we refer to this phase as the density-wave--spin-wave (DW-SW) state. Recall that $|\mathcal{N}_\downarrow|=|\mathcal{N}_\uparrow|$ due to the symmetric choice of parameters. The amplitude of the density modulations increases with increasing $\eta$, indicated by the increasing density-wave order parameters $|\mathcal{N}_\tau|$. Because of the direct pumping of two of the cavity modes and the presence of two-photon Raman processes which scatter photons into the unpumped modes without transfering momentum to the atoms as shown in~\fref{fig:processes}(c), there is no threshold behavior for the onset of the density waves. This is in contrast to transversally pumped ring cavities~\cite{nagy2006self,Mivehvar2018} and to the case where both pairs of modes $\{\hat{a}_\pm,\hat{b}_\pm\}$ couple to the same atomic transition~\cite{ostermann2015atomic}. The angle in the $x$-$y$ plane which is sweeped by the local pseudospin vector $\mathbf{s}(z)$ over a distance of $\lambda/2$, is always restricted to the interval $[0,\pi/2]$ for the entire parameter regime of the DW-SW phase. Hence, the spin-wave oscillations remain small in this phase. A typical spin-wave texture in the DW-SW phase is illustrated in~\fref{fig:spin}(a), where the local pseudospin vector $\mathbf{s}(z)$ exhibits small oscillations in the $x$-$y$ plane. The change in the length of the local pseudospin vector is due to the density modulations of the condensates. Note that the local pseudospin vector $\mathbf{s}(z)$ always lies in the $x$-$y$ plane, as $s_z(z)=0$ owing to the symmetric choice of the parameters. 

The second phase, corresponding to the uniform black region in the phase diagram of~\fref{fig:phase_diag}, is the plane-wave--spin-spiral (PW-SS) state. In this regime the density-wave order parameters are identically zero, $|\mathcal{N}_\tau|=0$, while the winding number is nonzero, $\mathcal{W}=1$. Therefore, the condensate densities are homogeneous in this phase, while the local pseudospin exhibits a spin-spiral texture. Figure~\ref{fig:spin}(b) depicts a representative spin-spiral state in this phase. As indicated by the winding number $\mathcal{W}=1$, the local pseudospin vector $\mathbf{s}(z)$ sweeps a full $2\pi$ angle over a $\lambda/2$ distance. The length of the local pseudospin vector is constant in space due to the uniform condensate densities in this phase. Note that again the local pseudospin vector has no $z$ component, $s_z(z)=0$.

In the third phase, both the density order parameters and the winding number are nonzero, $\{|\mathcal{N}_\tau|\neq0,\mathcal{W}=1\}$. This implies that the density wave and the spin-spiral coexist in this phase, hence the name density-wave--spin-spiral (DW-SS) state. The BEC densities exhibit strong modulations in this phase. The local pseudospin vector $\mathbf{s}(z)$, therefore, inherits this and its length changes drastically in space, while again sweeping a full $2\pi$ angle over a $\lambda/2$ distance as shown in Fig.~\ref{fig:spin}(c). As for the other two phases, the local pseudospin vector lies in the $x$-$y$ plane in the DW-SS phase.

For small cavity detuning, $\Delta\lesssim-9\omega_{\rm rec}$, the system becomes unstable (hashed region in Fig.~\ref{fig:phase_diag}). This is due to the fact that the effective relative cavity detuning, i.e., the dispersively shifted bare relative cavity detuning due to the presence of the atoms, becomes positive (i.e., blue detuned) resulting in cavity heating. In contrast to a single component BEC in a cavity where the dispersive shift is solely given by the factor $U_0 N$, an analytical expression for the dispersive shift of the cavity detuning in our model is nontrivial due to the various dispersive terms and coupling terms in the effective cavity-field Hamiltonian density~\eqref{eq:app-dynamic_matrix}.

The collective atomic spins $S_\pm$ and the spin-wave order parameters $\mathcal{S}_\pm^{(\pm)}$ exhibit distinct behaviors in these three phases. Figures~\ref{fig:spin-comp-vs-eta}(a) and \ref{fig:spin-comp-vs-eta}(b) show the absolute values of $S_\pm$ (solid blue curves) and $\mathcal{S}_\pm^{(\pm)}$ (dashed red curves) as a function of the rescaled pump strength $\sqrt{N}\eta/\omega_{\rm rec}$ for constant cavity detunings $\Delta/ \omega_\mathrm{rec}=-10$ and $-20$, respectively. By increasing the pump strength $\eta$ from zero, for $\Delta=-10 \omega_\mathrm{rec}$ in Fig.~\ref{fig:spin-comp-vs-eta}(a) the system undergoes a phase transition from the DW-SW state to the DW-SS state, while for $\Delta=-20 \omega_\mathrm{rec}$ in Fig.~\ref{fig:spin-comp-vs-eta}(b) the phase transition from the DW-SW to the DW-SS occurs indirectly via the intermediate PW-SS state (cf.~Fig.~\ref{fig:phase_diag}). While the collective atomic spins $S_\pm$ are nonzero in both DW-SW and DW-SS phases, it vanishes in the PW-SS phase. The latter can be understood by the fact that the local pseudospin vector $\mathbf{s}(z)$ has a constant length over space in the PW-SS phase and it does a full $2\pi$ rotation uniformly over one unit cell, resulting in $S_\pm=\int s_x(z) dz \pm \int is_y(z) dz=0$. On the other hand the spin-wave order parameters $\mathcal{S}_\pm^{(\pm)}$ are nonzero in all three phases, indicating spin modulations in all regimes.

The winding number $\mathcal{W}$ jumps from zero to one across the phase transitions from the DW-SW to the PW-SS/DW-SS, signaling that these are topological phase transitions (the red solid curve in the phase diagram in \fref{fig:phase_diag}, which is somewhat ragged due to the extremely slow convergence of numerics around these phase boundaries). In addition, the density-wave order parameters $\mathcal{N}_\tau$, the collective atomic spins $S_\pm$, and the spin-wave order parameters $\mathcal{S}_\pm^{(\pm)}$ exhibit discontinuous behaviors on the onset of these phase transitions (see Figs.~\ref{fig:phase_diag} and \ref{fig:spin-comp-vs-eta}). This indicates that the topological phase transitions from the DW-SW state to the PW-SS/DW-SS states also have first-order characteristics. 

Although the atomic parameters $\{\mathcal{N}_\tau,S_\pm,\mathcal{S}_\pm^{(\pm)}\}$ change continuously across the PW-SS to DW-SS phase transition [see Figs.~\ref{fig:phase_diag} and \ref{fig:spin-comp-vs-eta}(b)], they exhibit nonanalytical behavior, a characteristic of a second-order phase transition. Therefore, the phase transition from the PW-SS to the DW-SS is second order (yellow dashed curve in the phase diagram in~\fref{fig:phase_diag}). Note that the winding number $\mathcal{W}$ is one in both phases, and therefore it does not change across this phase transition. 

The phase boundary between the DW-SW and PW-SS phases is linear, whereas the other phase boundaries show more complex behaviors. All phase boundaries with different natures (i.e., topological first-order, and topologically trivial second-order phase transitions) meet at a single tricritical point, denoted by a green dot in the phase diagram in Fig.~\ref{fig:phase_diag}.
\begin{figure*}
\centering
(a) \includegraphics[width=0.45\textwidth]{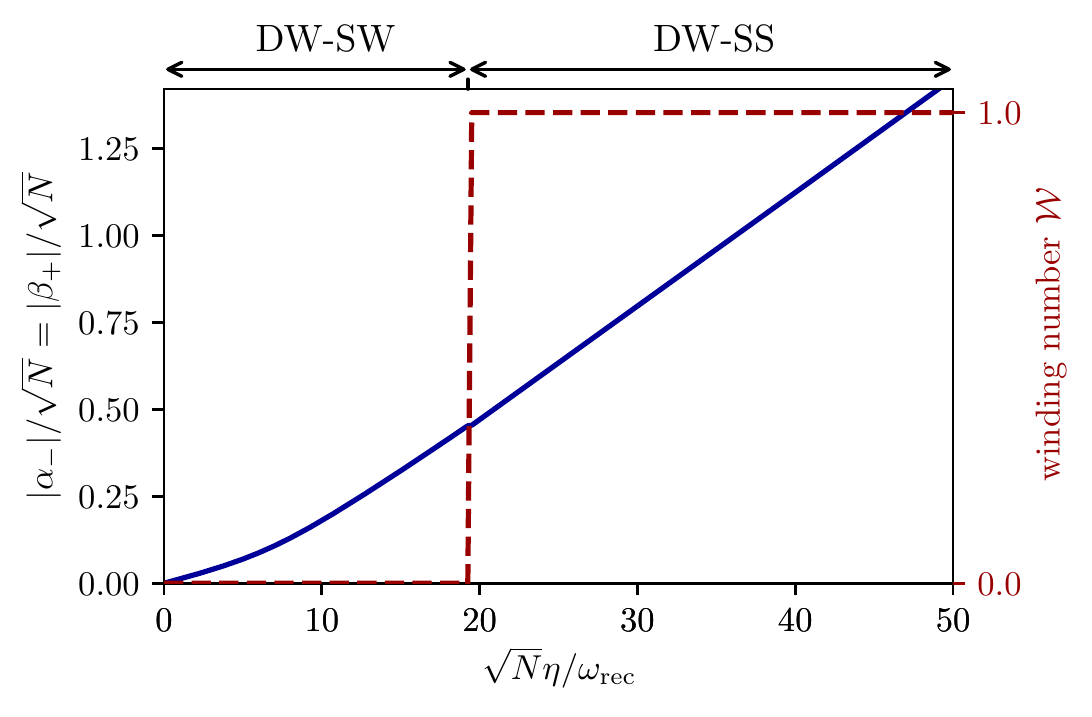}
(b) \includegraphics[width=0.45\textwidth]{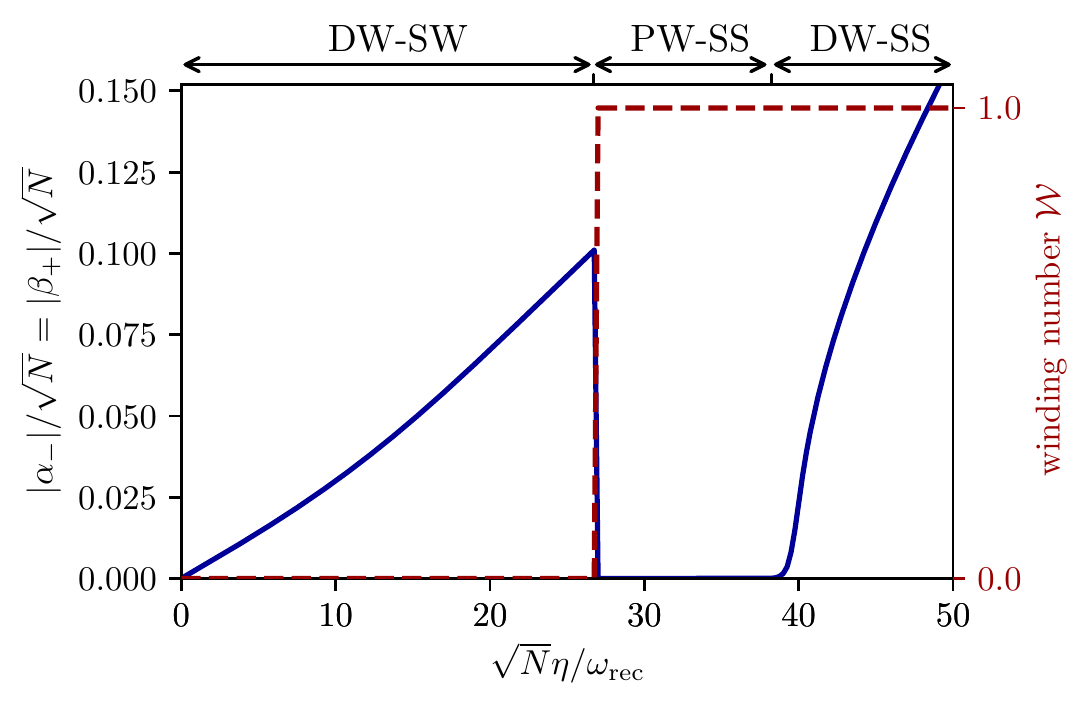}
\caption{
The absolute values of the cavity-field order parameters $|\alpha_-|/\sqrt{N}=|\beta_-|/\sqrt{N}$ (blue solid curve) an the winding number $\mathcal{W}$ (red dashed curve) as a function 
of the rescaled pump strength $\sqrt{N}\eta/\omega_{\rm rec}$ for $\Delta=-10\omega_{\rm rec}$~(a)
and $\Delta=-20\omega_{\rm rec}$~(b).
The cavity-field order parameters display first-order (second-order) characteristics 
across the phase transitions from the DW-SW to the PW-SS/DW-SS 
(from the PW-SS to the DW-SS). The first-order phase transition coincides
with the topological phase transition, where the winding number $\mathcal{W}$ jumps from zero to one.  
The other parameters are the same as \fref{fig:phase_diag}.}
\label{fig:cavity_topo_trans}
\end{figure*}

\subsection{Cavity-field phase diagram}
\label{subsec:fields}

The mean-field amplitudes of the unpumped cavity modes $\alpha_-$ and $\beta_+$ can be exploited as the cavity-field order parameters to further characterize the system. The inset of~\fref{fig:phase_diag} shows the absolute values of the rescaled unpumped modes $|\alpha_-|/\sqrt{N}=|\beta_+|/\sqrt{N}$ in the rescaled parameter space $\{\sqrt{N}\eta/\omega_{\rm rec},\Delta/\omega_{\rm rec}\}$. The absolute values of the unpumped modes are equal to each other once again due to the symmetric choice of the parameters. The cavity-field phase diagram has a similar form as the atomic phase diagram. In particular, the field phase boundaries coincide precisely with the atomic phase boundaries. The field order parameters $\{\alpha_-,\beta_+\}$ are nonzero in the DW-SW and DW-SS phases and increase monotonically by increasing pump strength. However, they are identically zero in the entire PW-SS phase. This can be understood by noting that the density-order parameters $\mathcal{N}_\tau$ and the collective spins $S_\pm$ are zero in this regime as discussed above. The dynamics of the two umpumped modes $\{\alpha_-,\beta_+\}$ then decouple completely from the pumped ones $\{\alpha_+,\beta_-\}$ [see Eq.~\eqref{eq:app-dynamic_matrix}], and, therefore, no photons are scattered into these unpumped modes in this phase. This leads to uniform potentials and Raman coupling [see Eqs.~\eqref{eq:pot-oper} and \eqref{eq:Raman-oper}], which in turn results in homogeneous condensate densities in a self-consistent manner. This signifies the nonlinear dynamical nature of the system.

Figures~\ref{fig:cavity_topo_trans}(a) and \ref{fig:cavity_topo_trans}(b) show cuts through the field phase diagram along the rescaled pump strength $\sqrt{N}\eta/\omega_{\rm rec}$ at constant cavity detunings $\Delta/ \omega_\mathrm{rec}=-10$ and $-20$, together with the corresponding winding numbers $\mathcal{W}$. The field-order parameters $\{\alpha_-,\beta_+\}$ exhibit similar behavior as the atomic parameters $\{\mathcal{N}_\tau,S_\pm,\mathcal{S}_\pm^{(\pm)}\}$. They display first-order (second-order) characteristics across the phase transitions from the DW-SW state to the PW-SS/DW-SS states (from the PW-SS state to the DW-SS phase), in accordance with the atomic phase transitions. Likewise, the first-oder phase transition coincides with the topological phase transition, where the winding $\mathcal{W}$ jumps from zero to one.  Therefore, there is a one to one correspondence between the atomic parameters $\{\mathcal{N}_\tau,S_\pm,\mathcal{S}_\pm^{(\pm)}\}$ on the one hand and the cavity-field order parameters $\{\alpha_-,\beta_+\}$ on the other hand. As a consequence, all the quantum phase transitions (and their natures) can be mapped out nondestructively through the cavity outputs. This is an important and distinct feature of the system.

\subsection{Atomic momentum distributions and cavity-induced spin-orbit coupling}
\label{subsec:SOC}

\begin{figure}[t!]
\centering
\includegraphics[width=0.45\textwidth]{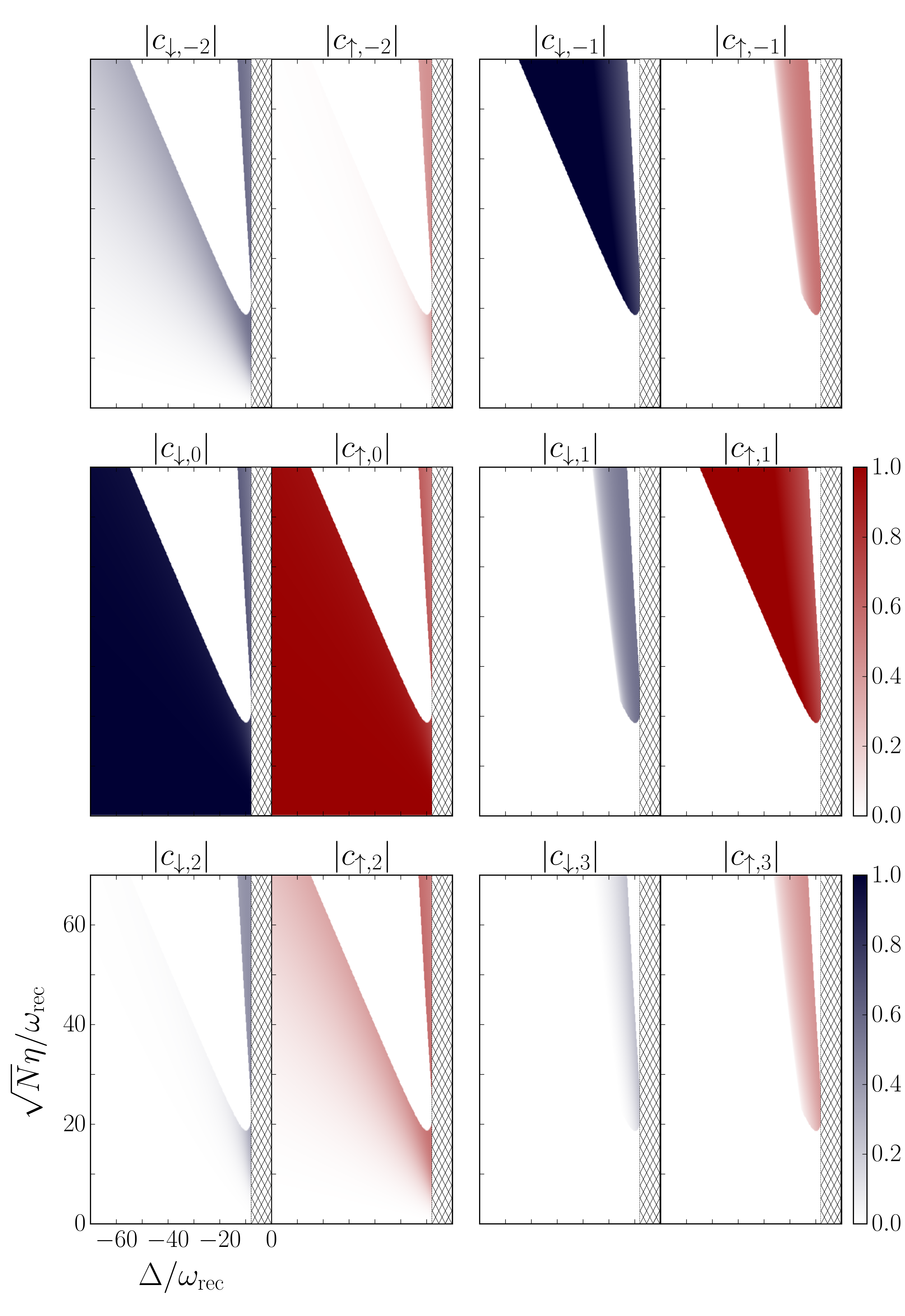}
\caption{The absolute values of the probability amplitudes $c_{\tau,j}$ of the lowest six momentum states $j\in\{0,\pm 1,\pm 2,+ 3\}$ for each condensate component $\tau$ in the parameter space $\{\sqrt{N}\eta/\omega_{\rm rec},\Delta/\omega_{\rm rec}\}$. For the the sake of clarity, vertical cuts of these diagrams for fixed $\Delta=-20\omega_{\rm rec}$ are presented in~\fref{fig:fourier_scan_delta_cut}. The other parameters are the same as \fref{fig:phase_diag}.}
\label{fig:fourier_scan}
\end{figure}

The discrete momentum exchange between the atoms and the light fields allows the decomposition of the condensate wave functions into plane waves $\psi_\tau(z)=\sum_{j=-\infty}^\infty c_{\tau,j}e^{ijkz}$. The absolute values of the probability amplitudes $c_{\tau,j}$ of the lowest six momentum states $j\in\{0,\pm 1,\pm 2,+ 3\}$ in the rescaled parameter space $\{\sqrt{N}\eta/\omega_{\rm rec},\Delta/\omega_{\rm rec}\}$ for each condensate component $\tau$ are shown in \fref{fig:fourier_scan}. Note that the even and odd momentum states do not coexist together. The boundary separating even and odd momenta coincides precisely with the topological phase boundary between the DW-SW state and the PW-SS/DW-SS states, illustrated in~\fref{fig:phase_diag}. The region where even (odd) momenta are occupied corresponds to the DW-SW (PW-SS or DW-SS) phase. In the DW-SW phase the zero momentum $c_{\tau,0}$ is the dominant state for both condensates and the nonzero higher momenta result in density modulations. In the PW-SS phase, the condensate wave functions are solely composed of one momentum component $|c_{\downarrow,1}|=|c_{\uparrow,-1}|=1/\sqrt{2}$, as expected for a homogeneous condensate. While in the DW-SS phase, higher odd momenta are also populated, leading to density modulations. For the sake of clarity, vertical cuts along the rescaled pump strength $\sqrt{N}\eta/\omega_{\rm rec}$  for constant cavity detuning  $\Delta=-20\omega_{\rm rec}$  of these momentum phase diagrams are also shown in Fig. 8.

\begin{figure}[t!]
\includegraphics[width=0.48\textwidth]{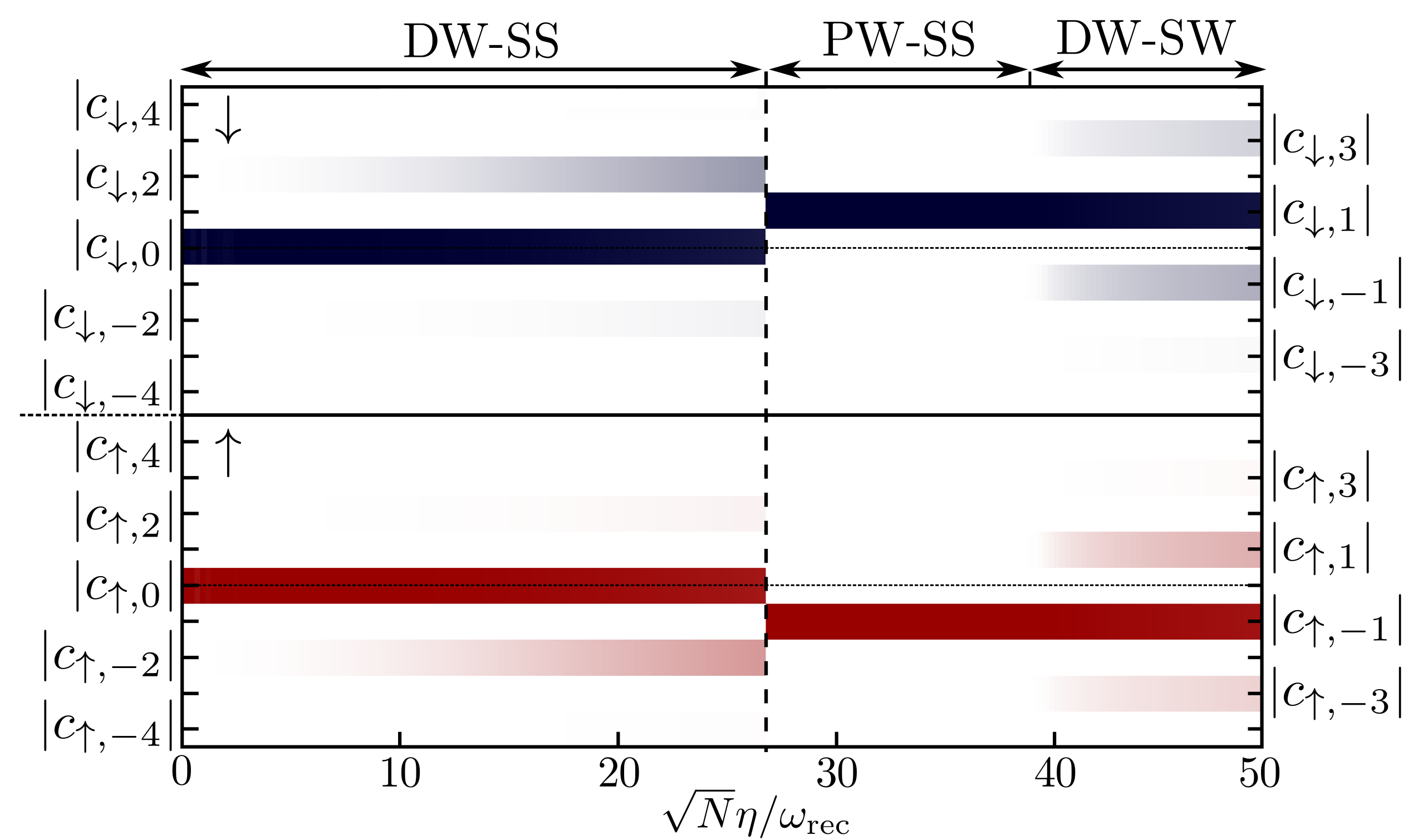}
\caption{The absolute values of the probability amplitudes $c_{\tau,j}$ of the lowest six momentum states $j\in\{0,\pm 1,\pm 2,+ 3\}$ for each condensate component $\tau$ as a function of $\sqrt{N}\eta/\omega_{\rm rec}$ for a constant $\Delta=-20\omega_{\rm rec}$. The even and odd momenta do not coexist. In the onset of the first-order topological phase transition from the DW-SW state to the PW-SS phase, the even momenta completely deplete and give way to the odd momenta. The other parameters are the same as \fref{fig:phase_diag}.}
\label{fig:fourier_scan_delta_cut}
\end{figure}

The phase boundaries where the even momenta completely deplete and the odd momenta are populated coincide exactly with the first-order topological phase transitions from the DW-SW state to the PW-SS/DW-SS state. That is, the onset of  the occupation of the odd momenta marks the appearance of spin spirals and the discrete jump of the winding number $\mathcal{W}$ from zero to one;
cf.\ Figs.~\ref{fig:spin-comp-vs-eta}(b), \ref{fig:cavity_topo_trans}(b) and \ref{fig:fourier_scan_delta_cut}.  This is intimately connected to the emergence of cavity-induced spin-orbit coupling for the atoms. The transition from the even to the odd momenta can be triggered and relaxed by (even very weak) two-body contact interactions, which are not included explicitly in our model.

The emergence of cavity-induced spin-orbit coupling can be most easily seen in the PW-SS phase, where the umpumped cavity modes $\{\alpha_-,\beta_+\}$ are zero. The effective atomic Hamiltonian density~\eqref{eq:eff-atomic-H-density} in the mean-field approximation then simplifies to   
\begin{align}
\mathcal{H}_{\rm SOC}&=\frac{1}{2m}(p I_{2\times2}-\hbar k\sigma_z)^2
+\frac{\hbar U_0}{2}(|\alpha_+|^2-|\beta_-|^2)\sigma_z\nonumber\\&
+\hbar \Omega_{0\rm R}\left(\alpha_+^*\beta_-\sigma_{\downarrow\uparrow}
+\alpha_+\beta^*_-\sigma_{\uparrow\downarrow}\right),
\label{eq:SOC_Ham}
\end{align}
after applying a unitary transformation~\cite{Mivehvar-2014}. Here $I_{2\times2}$ is the identity matrix in the pseudospin space, $\sigma_z$ is the third Pauli matrix and $\sigma_{\downarrow\uparrow}$ and $\sigma_{\uparrow\downarrow}$ are the transition matrices in the pseudospin basis. The Hamiltonian~\eqref{eq:SOC_Ham} has exactly the form of an equal Rashba-Dresselhaus spin-orbit coupled Hamiltonian, saving that the the Raman coupling now depends on the cavity fields $\{\alpha_+,\beta_-\}$ and is determined self-consistently. This Hamiltonian has been studied before in Refs.~\cite{Mivehvar-2014,Dong-2014,Mivehvar-2015} and indeed exhibits characteristics of spin-orbit coupled quantum gases, with extra features resulting from the dynamical nature of the synthetic spin-orbit coupling.

The effect of the spin-orbit coupling can be seen in the momentum distributions of the condensate wavefuctions in the PW-SS phase, where different pseudospin states are coupled to different momentum states. That is, the pseudospin down is solely coupled to the $+\hbar k$ momentum (recall that in the PW-SS phase $|c_{\downarrow,1}|=1/\sqrt{2}$), while the the pseudospin up is only coupled to the $-\hbar k$ momentum ($|c_{\uparrow,-1}|=1/\sqrt{2}$). Since the PW-SS phase sets in at large pump strengths $\eta$, the effective Raman transition rate $\Omega_{0\rm R}\alpha_+^*\beta_-$ is, therefore, always large. Hence, the single-particle energy dispersion of the spin-orbit coupled Hamiltonian~\eqref{eq:SOC_Ham} possesses a single minimum at $p=0$, as expected for large Raman transition rates~\cite{lin2011spin}. This is also the reason that both condensates have equal particle numbers, $|c_{\downarrow,1}|=|c_{\uparrow,-1}|=1/\sqrt{2}$, as the state at $p=0$ has an equal contribution from the up and down components due to the symmetric choice of the parameters.

Despite of the fact that in the DW-SS phase the effective atomic Hamiltonian density~\eqref{eq:eff-atomic-H-density} cannot be recast in the usual form of the equal Rashba-Dresselhaus spin-orbit-coupled Hamiltonian~\eqref{eq:SOC_Ham}, the cavity-induced synthetic spin-orbit coupling still manifests itself in the momentum distributions of the condensate wave functions. Although higher odd momenta are also populated, the $\pm\hbar k$ momenta are still the dominant states, and different pseudospin states are strongly coupled to solely one of them, $|c_{\downarrow,1}|=|c_{\uparrow,-1}|\lesssim1/\sqrt{2}$. This is in sharp contrast to the DW-SW phase, where both pseudospin states couple to the same zero momentum state, $|c_{\downarrow,0}|=|c_{\uparrow,0}|\lesssim1/\sqrt{2}$, resulting in no cavity-induced spin-orbit coupling. 

The period doubling of the condensate wave functions in the spin-orbit coupled PW-SS and DW-SS regimes, as it was mentioned at the beginning of Sec.~\ref{sec:phase_diagram}, can be understood through the momentum decomposition of the wave functions in these phases. In these spin-orbit coupled states, the condensate wave functions $\psi_\tau(z)=\sum_{j=-\infty}^\infty c_{\tau,2j+1}e^{i(2j+1)kz}$ are composed of solely odd momenta and are, therefore, $\lambda$ periodic. This is in contrast to the $\lambda/2$ periodicity of the condensate wave functions $\psi_\tau(z)=\sum_{j=-\infty}^\infty c_{\tau,2j}e^{2ijkz}$ in the DW-SW state (which are comprised of only even momenta) and the Hamiltonian density~\eqref{eq:eff-atomic-H-density}. 

The unpumped cavity modes $\{\hat{a}_-,\hat{b}_+\}$ play an important role in the emergence of cavity-induced spin-orbit coupling beyond a threshold for the pump-strength. This can be understood by re-examining the possible spin flipping processes in the system (see~\fref{fig:processes}). The essential photon-scattering processes for the spin-orbit coupling are the ones depicted in~\fref{fig:processes}(d), where photons are scattered between the modes $\hat{a}_\pm\leftrightarrow\hat{b}_\mp$ (via the atomic pseudospin flipping $\ket\downarrow\leftrightarrow\ket\uparrow$) and a $\pm2\hbar k$ momentum is transferred to the atom. Whereas the scattering processes shown in~\fref{fig:processes}(c), where photons are scattered between the modes $\hat{a}_\pm\leftrightarrow\hat{b}_\pm$ (again via the atomic pseudospin flipping $\ket\downarrow\leftrightarrow\ket\uparrow$) without any momentum kick to the atom, are not vital for the spin-orbit coupling. However, triggering the former processes costs more energy than the latter ones due to the atomic kinetic energy gain. Therefore, for lower pump strengths (i.e., in the DW-SW phase) the spin flipping processes with no momentum kick to the atoms are energetically favored and are the dominant processes. The essential spin-orbit coupling processes become energetically favored and dominant beyond the pump-strength threshold on the onset of the PW-SS and DW-SS states, where the sum of the kinetic energies of the odd momenta becomes less than the corresponding even ones.

The interplay between spin flipping processes with and without momentum transfer to the atom can be seen by comparing the collective atomic spins $S_\pm$ and  the spin-wave order parameters $\mathcal{S}_\pm^{(\pm)}$. Recall that $S_\pm$ ($\mathcal{S}_\pm^{(\pm)}$) quantifies the spin flipping processes without (with $\pm2\hbar k$) momentum kick to the atom. As can be seen from \fref{fig:spin-comp-vs-eta}, the collective spins $S_\pm$ are the dominant quantities in the DW-SW phase, while the spin-wave order parameters $\mathcal{S}_\pm^{(\pm)}$ become dominant only in the PW-SS and DW-SS phases. Consequently, cavity-induced spin-orbit coupling emerges only in the PW-SS and DW-SS regimes. This is in sharp contrast to the free space spin-orbit-coupled BEC, where spin-orbit coupling emerges at an infinitesimal Raman frequency~\cite{lin2011spin}.

\section{Collective Excitations} 
\label{sec:coll_ex}

In order to check the stability of our mean-field results and to obtain a deeper understanding of the system, we calculate the collective excitations of the system above the mean-field steady-states. To this end, we linearize the Heisenberg equations of motion~\eqref{eq:app-coupled_eff_Heisenberg_eqs} for quantum fluctuations of both atomic condensate wave functions $\delta\psi_\tau(x,t)=\delta\psi_\tau^{(+)}(x) e^{-i\omega t}+[\delta\psi_\tau^{(-)}(x)]^* e^{i\omega^* t}$, and field mode fluctuations $\delta\alpha_\pm(t)=\delta\alpha_\pm^{(+)} e^{-i\omega t}+[\delta\alpha_\pm^{(-)}]^* e^{i\omega^* t}$ and $\delta\beta_\pm(t)=\delta\beta_\pm^{(+)} e^{-i\omega t}+[\delta\beta_\pm^{(-)}]^* e^{i\omega^* t}$ around the mean-field stationary solutions $\{\psi_{0\tau}(x),\alpha_{0\pm},\beta_{0\pm}\}$. The linearized equations can be recast in matrix form,
\begin{equation}\label{eq:Bogoliubov_eq}
\omega\mathbf{f}=\mathbf{M}_{\rm B}\mathbf{f}
\end{equation}
where $\mathbf{f}$ is a vector composed of the atomic condensate and the field-mode fluctuations $\{\delta\psi_\tau^{(\pm)},\delta\alpha_\pm^{(\pm)},\delta\beta_\pm^{(\pm)}\}$ and $\mathbf{M}_{\rm B}$ is a (nonhermitian) Bogoliubov matrix. We relegate the details to Appendix~\ref{app:B}.

\begin{figure}[t!]
\centering
\includegraphics[width=0.47\textwidth]{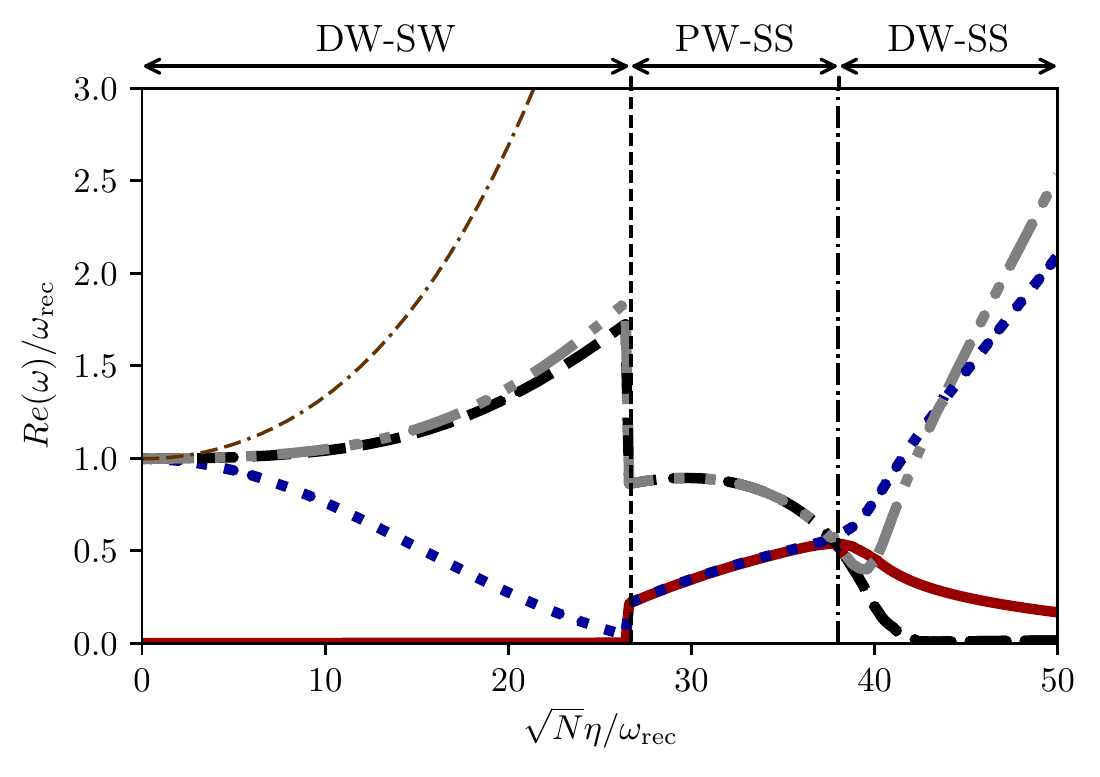}
\caption{Real part of the five lowest-lying collective excitations of the system for $\Delta=-20\omega_\mathrm{rec}$
as a function of the rescaled pump strength $\sqrt{N}\eta/\omega_{\rm rec}$. 
The other parameters are the same as \fref{fig:phase_diag}.}
\label{fig:coll_ex}
\end{figure}

We numerically diagonalize the Bogoliubov matrix $\mathbf{M}_{\rm B}$ as a function of the pump strength to obtain the collective excitation spectrum $\omega(\eta)$. Figure~\ref{fig:coll_ex} shows the real part of the five lowest lying positive-frequency excitations of the system as a function of the rescaled pump strength $\sqrt{N}\eta/\omega_\mathrm{rec}$ at a fixed cavity detuning $\Delta=-20\omega_\mathrm{rec}$.

In the DW-SW regime corresponding to $\sqrt{N}\eta\lesssim27\omega_\mathrm{rec}$, there exists a gapless Goldstone mode, \ie
$\text{Re}(\omega)=0$. This is associated with the spontaneously broken continuous screw-like symmetry of the system in this regime. This symmetry breaking is a consequence of the occupation of the even (in particular, the zero) momentum states in this phase, which leads to wave functions of the form $\psi_\tau(z)=(c_{\tau,0}+c_{\tau,\pm 2}e^{\pm 2ikz}+\dots)$. A spatial translation of the wave functions $\psi_\tau(z)\rightarrow \psi_\tau(z+\Delta z)$ cannot be compensated by the phase rotation of the condensate wave functions $\psi_\downarrow\rightarrow \psi_\downarrow e^{-ik\Delta z}$ and $\psi_\uparrow\rightarrow\psi_\uparrow e^{ik\Delta z}$, because of the occupation of the zero momentum state $c_{\tau,0}\neq0$. Hence, the condensate wave functions in the DW-SW regime are not invariant under the continuous symmetry group of the Hamiltonian and, therefore, they spontaneously break the continuous symmetry of the system. The appearance of this extra gapless Goldstone mode, in addition to the gapless phonon mode resulting from the spontaneous breaking of the internal gauge symmetry which is not shown here, indicates that the DW-SW state is a supersolid, \ie a state with two spontaneously broken continuous symmetries~\cite{leonard2017supersolid,Li2017stripe,Mivehvar2018}. A supersolid has the characteristics of both a crystal and a superfluid, that is, it is a state with a long-range periodic density order which can flow without experiencing any friction force.

For infinitely small values of $\eta$, the collective excitations at frequencies $\sim\omega_\mathrm{rec}$ are fourfold degenerate. These excitations correspond to condensate fluctuations of both BEC components at momenta $\pm\hbar k$. With increasing $\eta$, one of these branches approaches the zero energy. The point where this excitation branch touches zero coincides exactly with the mean-field critical pump strength for the phase transition from the DW-SW state to the PW-SS phase (dashed line in~\fref{fig:coll_ex}). As soon as the gap of the lowest $\pm\hbar k$ branch closes, the odd momenta become the lower energy states and the synthetic spin-orbit coupling emerges. Hence, it is the interplay between the two lowest even and odd excitation branches (solid red and dotted blue in~\fref{fig:coll_ex}) that leads to emergent spin-orbit coupling in this system.

In the PW-SS regime all excitation branches are doubly degenerate and gapped, indicating that the continuous symmetry of the system is not broken. This is intuitively obvious since the spin spiral with no density modulation perfectly respects the screw-like symmetry of the system. This can be seen most readily by the fact that the condensate wave functions $\psi_\downarrow(z)=c_{\downarrow,1}e^{ik z}$ and $\psi_\uparrow(z)=c_{\uparrow,-1}e^{-ik z}$ are invariant under the screw-like symmetry transformation: $\psi_\downarrow(z)\rightarrow \psi_\downarrow(z+\Delta z)e^{-ik\Delta z}=\psi_\downarrow(z)$ and $\psi_\uparrow(z)\rightarrow \psi_\uparrow(z+\Delta z)e^{ik\Delta z}=\psi_\uparrow(z)$.

With further increasing pump strength $\eta$ the degenerate branches start to split up at the value of $\eta$ which perfectly coincides with the mean-field critical pump strength for the phase transition from the PW-SS state to the DW-SS regime (horizontal dash dotted line). One of the excitation branches (dashed black curve) exhibits an exotic ``quasi" gapless-mode behavior in large pump strengths. This can be understood by examining the condensate wave functions in this regime. For the sake of simplicity we restrict our argument to one pseudospin, say, $\psi_\downarrow(z)=(c_{\downarrow,1}e^{ikz}+c_{\downarrow,-1}e^{-ik z}+c_{\downarrow,3}e^{3ikz}+\dots$). In general, this state is not invariant under the screw-like symmetry transformation $\psi_\downarrow(z)\rightarrow \psi_\downarrow(z+\Delta z)e^{-ik\Delta z}\neq\psi_\downarrow(z)$. That said, at the onset of the DW-SS state $c_{\downarrow,1}$ is the dominant probability amplitude in this expansion as can be seen from~\fref{fig:fourier_scan}, i.e., $|c_{\downarrow,1}|\gg\{|c_{\downarrow,-1}|,|c_{\downarrow,3}|,\cdots\}$. Therefore, this wave function can be approximated as $\psi_\downarrow(z)\simeq c_{\downarrow,1}e^{ik z}$, which approximately preserves the screw-like symmetry of the system. For larger pump strengths, however, the higher momentum-state coefficients are not negligible anymore, resulting in a state which breaks the screw-like symmetry of the system. This leads to the appearance of a quasi gapless mode~$\sim0\omega_\mathrm{rec}$ at large $\eta$, which never reaches the zero energy exactly because $c_{\downarrow,1}$ remains dominant throughout the entire regime. This is in stark contrast to the DW-SW state, where even in the onset of this phase at very small pump strengths the screw-like symmetry is broken completely. This is due to the dominant population of the zero momentum state $|c_{\tau,0}|\gg\{|c_{\tau,\pm2}|,|c_{\tau,\pm4}|,\cdots\}$ in this entire regime (recall that the zero momentum state is the one that breaks the screw-like symmetry of the system in the DW-SW state).  

It should be mentioned that imaginary parts of the collective excitations remain remarkably small for a dissipative system, \ie $\kappa\neq0$. This is a hint for the dynamical stability of the phases which is also confirmed by performing a real time evolution of the stationary state solutions.

\section{Conclusion and Outlook}
\label{sec:conclusion_and_outlook}

We theoretically studied an effective two-component BEC inside a ring cavity, which possesses two pairs of nearly resonant running-wave modes with orthogonal polarizations. Our proposed model takes into account both atomic internal and external degrees of freedom, as well as the field amplitude and polarizations degrees of freedom. We predict that even in the simplest symmetric choice of parameters, the interplay between various degrees of freedom already results in novel phases and exotic quantum phase transitions of different natures. All the phases and the quantum phase transitions between them can be readily realized by solely tuning the frequencies and powers of the pump lasers, relevant parameters in cavity-QED experiments~\cite{Schmidt-2014,Landig-2016,schuster2018pinning}. Remarkably, all the quantum phase transitions, including the topological one, can be monitored directly through the cavity outputs. Our proposal can be implemented with minor modifications to state-of-the-art experiments in cavity QED~\cite{Kruse2003,Nagorny2003,Slama-2007,Zimmermann2007,Bux2013,Schmidt-2014,naik2017bec, schuster2018pinning} and it may open a new direction for studying topological effects in ultracold atoms via \textit{in situ} monitoring. Additional physics may arise for asymmetric choices of the parameters as well as the inclusion of large two-body contact interactions. However, we leave the investigation of these interesting issues for future works.

\begin{acknowledgements}
We thank C. Simon, F. Piazza, and C. Zimmermann for fruitful discussions. We acknowledge support by the Austrian Science Fund FWF through the Projects SFB FoQuS P13 and No. I1697-N27.

\end{acknowledgements}

\appendix
\section{Adiabatic Elimination of the excited state}
\label{app:A}

Here we demonstrate how the model given in Eq.~\eqref{eq:app-coupled_eff_Heisenberg_eqs} can be obtained from the single-particle Hamiltonian density~\eqref{eq:total-single-atom-H-density}.
The single-particle Hamiltonian density~\eqref{eq:total-single-atom-H-density} can be transferred into the rotating frame of the pump lasers through $\tilde{\mathcal{H}}=\mathcal{U}\mathcal{H}\mathcal{U}^\dag+i\hbar(\partial_t\mathcal{U})\mathcal{U}^\dag$ and exploiting the unitary transformation 
\begin{align*} 
\mathcal{U}=\exp \Big\{i
\Big[(\hat{a}_+^\dag\hat{a}_++\hat{a}_-^\dag\hat{a}_--\hat{\sigma}_{\downarrow\downarrow})\tilde\omega_a\nonumber\\
+(\hat{b}_+^\dag\hat{b}_++\hat{b}_-^\dag\hat{b}_--\hat{\sigma}_{\uparrow\uparrow})\tilde\omega_b\Big]t
\Big\}.
\end{align*}    
The corresponding many-body Hamiltonian expressed in the formalism of second quantization then reads,
\begin{align} \label{eq:totall-H}
H&=\int dz \hat{\boldsymbol\Psi}^\dag 
{\mathcal M}\hat{\boldsymbol\Psi}
-\hbar\sum_{j=+,-} 
\left(\Delta_a \hat{a}^\dag_j \hat{a}_j+\Delta_b\hat{b}^\dag_j \hat{b}_j\right)\nonumber\\
&+i\hbar\left[\eta_+ \hat{a}_+^\dag+\eta_- \hat{b}_-^\dag- \text{H.c.}\right],
\end{align}
where 
$\hat{\boldsymbol\Psi}(z)=(\hat{\psi}_\downarrow(z),\hat{\psi}_\uparrow(z),\hat{\psi}_e(z))^\mathsf{T}$, and
\begin{align}
{\mathcal M}=
\begin{pmatrix}
\frac{\hat{p}^2}{2m}-\frac{\hbar\delta}{2} & 0 & \mathcal{M}_{13} \\
0 & \frac{\hat{p}^2}{2m}+\frac{\hbar\delta}{2} & \mathcal{M}_{23} \\
\mathcal{M}_{31} & \mathcal{M}_{32} & \frac{\hat{p}^2}{2m}-\frac{\hbar}{2}(\Delta_\downarrow+\Delta_\uparrow)
\end{pmatrix},
\end{align}
with the elements
\begin{align}
\mathcal{M}_{31}=\mathcal{M}_{13}^\dag=
\hbar \mathcal{G}_\downarrow\left(e^{ikz}\hat{a}_++e^{-ikz}\hat{a}_-\right),\nonumber\\
\mathcal{M}_{32}=\mathcal{M}_{23}^\dag=
\hbar \mathcal{G}_\uparrow\left(e^{ikz}\hat{b}_++e^{-ikz}\hat{b}_-\right).
\end{align}
The constant term $[(\omega_\downarrow+\tilde\omega_a)/2+(\omega_\uparrow+\tilde\omega_b)/2]I_{3\times3}$ is omitted.
The dynamics of the atomic and cavity field operators can be determined by simultaneously solving the following Heisenberg equations of motion $i\hbar\partial_t\hat\psi_\tau=[\psi_\tau,H]$ and $i\hbar\partial_t\hat{a}_j/\hat{b}_j=[\hat{a}_j/\hat{b}_j,H]-i\hbar\kappa\hat{a}_j/\hat{b}_j$. Substituting the Hamiltonian~\eqref{eq:totall-H} leads to the following set of coupled differential equations
\begin{align}\label{eq:heisenberg_eq_motion}
i\hbar\frac{\partial}{\partial t}\hat\psi_\downarrow&=
\left(\frac{\hat{p}^2}{2m}-\frac{\hbar}{2}\delta\right)\hat\psi_\downarrow
+\hbar \mathcal{G}_\downarrow^*\left(e^{-ikz}\hat{a}_+^\dag+e^{ikz}\hat{a}_-^\dag\right)\hat\psi_e,
\nonumber\\
i\hbar\frac{\partial}{\partial t}\hat\psi_\uparrow&=
\left(\frac{\hat{p}^2}{2m}+\frac{\hbar}{2}\delta\right)\hat\psi_\uparrow
+\hbar \mathcal{G}_\uparrow^*\left(e^{-ikz}\hat{b}_+^\dag+e^{ikz}\hat{b}_-^\dag\right)\hat\psi_e,
\nonumber\\
i\hbar\frac{\partial}{\partial t}\hat{a}_+&=-\hbar(\Delta_a+i\kappa)\hat{a}_+
+\hbar \mathcal{G}_\downarrow^*\int dz e^{-ikz} \hat\psi_\downarrow^\dag\hat\psi_e+i\hbar\eta_+,
\nonumber\\
i\hbar\frac{\partial}{\partial t}\hat{a}_-&=-\hbar(\Delta_a+i\kappa)\hat{a}_-
+\hbar \mathcal{G}_\downarrow^*\int dz e^{ikz} \hat\psi_\downarrow^\dag\hat\psi_e,
\nonumber\\
i\hbar\frac{\partial}{\partial t}\hat{b}_+&=-\hbar(\Delta_b+i\kappa)\hat{b}_+
+\hbar \mathcal{G}_\uparrow^*\int dz e^{-ikz} \hat\psi_\uparrow^\dag\hat\psi_e,
\nonumber\\
i\hbar\frac{\partial}{\partial t}\hat{b}_-&=-\hbar(\Delta_b+i\kappa)\hat{b}_-
+\hbar \mathcal{G}_\uparrow^*\int dz e^{ikz} \hat\psi_\uparrow^\dag\hat\psi_e+i\hbar\eta_-.
\end{align}
If the relative atomic detunings $\Delta_{\uparrow}$ and $\Delta_{\downarrow}$ are large compared to the two-photon detuning $\delta$ and the atom-photon couplings $\mathcal{G}_\uparrow$ and $\mathcal{G}_\downarrow$, the atomic excited state reaches a steady-state on a short time scale and its dynamics can be eliminated adiabatically. By setting $\partial_t\hat\psi_e=0$ in the Heisenberg equation of motion for $\hat\psi_e$ and omitting the kinetic energy compared to the term proportional to $\hbar(\Delta_{\uparrow}+\Delta_{\downarrow})$, we obtain the steady-state field operator of the atomic excited state
\begin{align} \label{eq:ss-atomic-e-field-op}
\hat\psi_e^{\rm ss}\simeq\frac{2}{(\Delta_\downarrow+\Delta_\uparrow)}
\Big[& \mathcal{G}_\downarrow\left(e^{ikz}\hat{a}_++e^{-ikz}\hat{a}_-\right)\hat\psi_\downarrow\nonumber\\
&+\mathcal{G}_\uparrow\left(e^{ikz}\hat{b}_++e^{-ikz}\hat{b}_-\right)\hat\psi_\uparrow\Big].
\end{align}
This steady-state field operator can be substituted into the  Heisenberg equations of motion~\eqref{eq:heisenberg_eq_motion} resulting in a set of six coupled nonlinear equations for $\{\hat\psi_\downarrow,\hat\psi_\uparrow,\hat{a}_\pm,\hat{b}_\pm\}$ given in~\eref{eq:app-coupled_eff_Heisenberg_eqs}.

\section{Linearized equations}
\label{app:B}

In this appendix we describe the calculations leading to the collective excitation spectrum presented in section~\ref{sec:coll_ex} in more detail. Plugging the ansatz $\psi_\tau(x,t)=\psi_{0\tau}(x)+\delta\psi(x,t)$, $\alpha_\pm(t)=\alpha_{\pm0}+\delta\alpha_\pm(t)$ and $\beta_\pm(t)=\beta_{\pm0}+\delta\beta_\pm(t)$ into the mean field version of~\eref{eq:app-coupled_eff_Heisenberg_eqs} and performing the Bogoliubov transformation as it is already discussed in section~\ref{sec:coll_ex} leads to the following linearized equations for the modes
\begin{widetext}
\small
\begin{align}\label{eq:lin_eq_modes}
\omega \delta\alpha_+^{(+)}&=-\tilde{\Delta}_a\delta\alpha_+^{(+)}+U_{0\downarrow} \mathcal{N}_\downarrow\delta\alpha_+^{(+)}+\Omega_{0\mathrm{R}} S_-\delta\beta_+^{(+)}+\Omega_{0\mathrm{R}} \mathcal{S}^{(1/2)}_-\delta\beta_-^{(+)}+U_{0\downarrow} \mathcal{A}_{+*}^\downarrow\delta\psi_\downarrow^{(+)}+U_{0\downarrow} \mathcal{A}_{+}^\downarrow\delta\psi_\downarrow^{(-)}+\Omega_{0\mathrm{R}}\mathcal{B}_{+*}^\downarrow\delta\psi_\uparrow^{(+)}+\Omega_{0\mathrm{R}}\mathcal{B}_{+}^\uparrow\delta\psi_\downarrow^{(-)},\nonumber\\
\omega \delta\alpha_+^{(-)}&=\tilde{\Delta}_a^*\delta\alpha_+^{(-)}-U_{0\downarrow} \mathcal{N}_\downarrow^*\delta\alpha_-^{(-)}-\Omega_{0\mathrm{R}}^* S_-^*\delta\beta_+^{(-)}-\Omega_{0\mathrm{R}}^* \mathcal{S}^{(1)*}_-\delta\beta_-^{(-)}-U_{0\downarrow} \mathcal{A}_{+*}^{\downarrow*}\delta\psi_\downarrow^{(-)}-U_{0\downarrow} \mathcal{A}_{+}^{\downarrow*}\delta\psi_\downarrow^{(+)}-\Omega_{0\mathrm{R}}^*\mathcal{B}_{+*}^{\downarrow*}\delta\psi_\uparrow^{(-)}-\Omega_{0\mathrm{R}}^*\mathcal{B}_{+}^{\uparrow*}\delta\psi_\downarrow^{(+)},\nonumber\\
\omega \delta\alpha_-^{(+)}&=-\tilde{\Delta}_a\delta\alpha_-^{(+)}+U_{0\downarrow} \mathcal{N}_\downarrow^*\delta\alpha_+^{(+)}+\Omega_{0\mathrm{R}} S_-\delta\beta_-^{(+)}+\Omega_{0\mathrm{R}} \mathcal{S}^{(2)}_-\delta\beta_+^{(+)}+U_{0\downarrow} \mathcal{A}_{-*}^\downarrow\delta\psi_\downarrow^{(+)}+U_{0\downarrow} \mathcal{A}_{-}^\downarrow\delta\psi_\downarrow^{(-)}+\Omega_{0\mathrm{R}}\mathcal{B}_{-*}^\downarrow\delta\psi_\uparrow^{(+)}+\Omega_{0\mathrm{R}}\mathcal{B}_{-}^\uparrow\delta\psi_\downarrow^{(-)},\nonumber\\
\omega \delta\alpha_-^{(-)}&=\tilde{\Delta}_a^*\delta\alpha_-^{(-)}-U_{0\downarrow} \mathcal{N}_\downarrow\delta\alpha_+^{(-)}-\Omega_{0\mathrm{R}}^* S_-^*\delta\beta_-^{(-)}-\Omega_{0\mathrm{R}}^* \mathcal{S}^{(2)*}_-\delta\beta_+^{(-)}-U_{0\downarrow} \mathcal{A}_{-*}^{\downarrow*}\delta\psi_\downarrow^{(-)}-U_{0\downarrow} \mathcal{A}_{-}^{\downarrow*}\delta\psi_\downarrow^{(+)}-\Omega_{0\mathrm{R}}^*\mathcal{B}_{-*}^{\downarrow*}\delta\psi_\uparrow^{(-)}-\Omega_{0\mathrm{R}}^*\mathcal{B}_{-}^{\uparrow*}\delta\psi_\downarrow^{(+)},\nonumber\\
\omega \delta\beta_+^{(+)}&=-\tilde{\Delta}_a\delta\beta_+^{(+)}+U_{0\uparrow} \mathcal{N}_\uparrow\delta\beta_-^{(+)}+\Omega_{0\mathrm{R}}^* S_-^*\delta\alpha_+^{(+)}+\Omega_{0\mathrm{R}} \mathcal{S}^{(2)*}_-\delta\alpha_-^{(+)}+U_{0\uparrow} \mathcal{B}_{+*}^\uparrow\delta\psi_\uparrow^{(+)}+U_{0\uparrow} \mathcal{B}_{+}^\uparrow\delta\psi_\uparrow^{(-)}+\Omega_{0\mathrm{R}}^*\mathcal{A}_{+*}^\uparrow\delta\psi_\downarrow^{(+)}+\Omega_{0\mathrm{R}}^*\mathcal{A}_{+}^\downarrow\delta\psi_\uparrow^{(-)},\nonumber\\
\omega \delta\beta_+^{(-)}&=\tilde{\Delta}_a^*\delta\beta_+^{(-)}-U_{0\uparrow} \mathcal{N}_\uparrow^*\delta\beta_-^{(-)}-\Omega_{0\mathrm{R}} S_-\delta\alpha_+^{(-)}-\Omega_{0\mathrm{R}} \mathcal{S}^{(2)}_-\delta\alpha_-^{(-)}-U_{0\uparrow} \mathcal{B}_{+*}^{\uparrow*}\delta\psi_\uparrow^{(-)}-U_{0\uparrow} \mathcal{B}_{+}^{\uparrow*}\delta\psi_\uparrow^{(+)}-\Omega_{0\mathrm{R}}\mathcal{A}_{+*}^{\uparrow*}\delta\psi_\downarrow^{(-)}-\Omega_{0\mathrm{R}}\mathcal{A}_{+}^{\downarrow*}\delta\psi_\uparrow^{(+)},\nonumber\\
\omega \delta\beta_-^{(+)}&=-\tilde{\Delta}_a\delta\beta_-^{(+)}+U_{0\uparrow} \mathcal{N}_\uparrow^*\delta\beta_+^{(+)}+\Omega_{0\mathrm{R}}^* S_-^*\delta\alpha_-^{(+)}+\Omega_{0\mathrm{R}}^* \mathcal{S}^{(1)*}_-\delta\alpha_+^{(+)}+U_{0\uparrow} \mathcal{B}_{-*}^\uparrow\delta\psi_\uparrow^{(+)}+U_{0\uparrow} \mathcal{B}_{-}^\uparrow\delta\psi_\uparrow^{(-)}+\Omega_{0\mathrm{R}}^*\mathcal{A}_{-*}^\uparrow\delta\psi_\downarrow^{(+)}+\Omega_{0\mathrm{R}}^*\mathcal{A}_{-}^\downarrow\delta\psi_\uparrow^{(-)},\nonumber\\
\omega \delta\beta_-^{(-)}&=\tilde{\Delta}_a^*\delta\beta_-^{(-)}-U_{0\uparrow} \mathcal{N}_\uparrow\delta\beta_+^{(-)}-\Omega_{0\mathrm{R}} S_-\delta\alpha_-^{(-)}-\Omega_{0\mathrm{R}} \mathcal{S}^{(1)}_-\delta\alpha_+^{(-)}-U_{0\uparrow} \mathcal{B}_{-*}^{\uparrow*}\delta\psi_\uparrow^{(-)}-U_{0\uparrow} \mathcal{B}_{-}^{\uparrow*}\delta\psi_\uparrow^{(+)}-\Omega_{0\mathrm{R}}\mathcal{A}_{-*}^{\uparrow*}\delta\psi_\downarrow^{(-)}-\Omega_{0\mathrm{R}}\mathcal{A}_{-}^{\downarrow*}\delta\psi_\uparrow^{(+)},
\end{align}
\end{widetext}
where we introduced the following shorthand notations
\begin{align}
\mathcal{A}_\pm^{\downarrow\uparrow}\xi:&=\int A_\pm^{\downarrow\uparrow}\xi dz,\nonumber \\
\mathcal{A}_{\pm*}^{\downarrow\uparrow}\xi:&=\int A_{\pm*}^{\downarrow\uparrow} \xi dz,\nonumber\\
\mathcal{B}_\pm^{\downarrow\uparrow}\xi:&=\int B_\pm^{\downarrow\uparrow} \xi dz,\nonumber\\
\mathcal{B}_{\pm*}^{\downarrow\uparrow}\xi:&=\int B_{\pm*}^{\downarrow\uparrow} \xi dz,
\end{align}
with
\begin{align}
A_\pm^{\downarrow\uparrow}&=\psi_0^{\downarrow\uparrow}\left(\alpha_{0\pm}+e^{\mp2 i kz}\alpha_{0\mp}\right),\nonumber\\
A_{\pm*}^{\downarrow\uparrow}&=\psi_0^{\downarrow\uparrow*}\left(\alpha_{0\pm}+e^{\mp2 i kz}\alpha_{0\mp}\right),\nonumber\\
B_\pm^{\downarrow\uparrow}&=\psi_0^{\downarrow\uparrow}\left(\beta_{0\pm}+e^{\mp2 i kz}\beta_{0\mp}\right),\nonumber\\
B_{\pm*}^{\downarrow\uparrow}&=\psi_0^{\downarrow\uparrow*}\left(\beta_{0\pm}+e^{\mp2 i kz}\beta_{0\mp}\right).
\end{align}
The linearized equations for the atomic degrees of freedom read in
\begin{widetext}
\begin{align}\label{eq:lin_eq_atoms}
\small
\omega \delta\psi_\downarrow^{(+)}&=\frac{1}{\hbar}\left[D_{\downarrow,1}-\mu\right]\delta\psi_\downarrow^{(+)}+\hbar \Omega_{\mathrm{R}}(z) \delta\psi_\uparrow^{(+)}+U_{0\downarrow}\left(A_{+*}^{\downarrow*}\delta \alpha_+^{(+)}+A_{-*}^{\downarrow*}\delta \alpha_-^{(+)}+A_+^\downarrow\delta \alpha_+^{(-)}+A_-^\downarrow\delta \alpha_-^{(-)}\right)\nonumber\\
&+\Omega_{0\mathrm{R}}\left(B_+^\uparrow\delta \alpha_+^{(-)}+B_-^\uparrow\delta \alpha_-^{(-)}+A_{+*}^{\uparrow*}\delta\beta_+^{(+)}+A_{-*}^{\uparrow*}\delta \beta_-^{(+)}\right),\nonumber\\
\omega \delta\psi_\downarrow^{(-)}&=-\frac{1}{\hbar}\left[D_{\downarrow,1}-\mu\right]^*\delta\psi_\downarrow^{(-)}-\hbar \Omega^*_{\mathrm{R}}(z)\delta\psi_\uparrow^{(-)}-U_{0\downarrow}\left(A_{+*}^\downarrow\delta \alpha_+^{(-)}+A_{-*}^\downarrow\delta \alpha_-^{(-)}+A_+^{\downarrow*}\delta \alpha_+^{(+)}+A_-^{\downarrow*}\delta \alpha_-^{(+)}\right)\nonumber\\
&-\Omega_{0\mathrm{R}}^*\left(B_+^{\uparrow*}\delta \alpha_+^{(+)}+B_-^{\uparrow*}\delta \alpha_-^{(+)}+A_{+*}^\uparrow\delta\beta_+^{(-)}+A_{-*}^\uparrow\delta \beta_-^{(-)}\right),\nonumber\\
\omega \delta\psi_\uparrow^{(+)}&=\frac{1}{\hbar}\left[D_{\uparrow,2}-\mu\right]\delta\psi_\uparrow^{(+)}+\hbar \Omega^*_{\mathrm{R}}(z)\delta\psi_\downarrow^{(+)}+U_{0\uparrow}\left(B_{+*}^{\uparrow*}\delta \beta_+^{(+)}+B_{-*}^{\uparrow*}\delta \beta_-^{(+)}+ B_+^\uparrow\delta \beta_+^{(-)}+B_-^\uparrow\delta \beta_-^{(-)}\right)\nonumber\\
&+\Omega_{0\mathrm{R}}^*\left(B_{+*}^{\downarrow*}\delta \alpha_+^{(+)}+B_{-*}^{\downarrow*}\delta \alpha_-^{(+)}+A_+^\downarrow\delta \beta_+^{(-)}+A_-^\downarrow\delta \beta_-^{(-)}\right),\nonumber\\
\omega \delta\psi_\uparrow^{(-)}&=-\frac{1}{\hbar}\left[D_{\uparrow,2}-\mu\right]^*\delta\psi_\uparrow^{(-)}-\hbar \Omega_{\mathrm{R}}(z)\delta\psi_\downarrow^{(-)}-U_{0\uparrow}\left(B_{+*}^\uparrow\delta \beta_+^{(-)}+B_{-*}^\uparrow\delta \beta_-^{(-)}+B_+^{\uparrow*}\delta \beta_+^{(+)}+ B_-^{\uparrow*}\delta \beta_-^{(+)}\right)\nonumber\\
&-\Omega_{0\mathrm{R}}\left(B_{+*}^\downarrow\delta \alpha_+^{(-)}+B_{-*}^\downarrow\delta \alpha_-^{(-)}+A_+^{\downarrow*}\delta \beta_+^{(+)}+ A_-^{\downarrow*}\delta \beta_-^{(+)}\right),
\end{align}
\end{widetext}
where we introduce the shorthand notation $D_{\downarrow\uparrow,i}:=-\frac{p^2}{2m}+\hbar U_{\downarrow\uparrow}(z)+(-1)^i\frac{\hbar\delta}{2} $ and $\mu$ denotes the chemical potential.

The set of equations~\eqref{eq:lin_eq_modes} and~\eqref{eq:lin_eq_atoms} can be written in matrix form which results in~\eref{eq:Bogoliubov_eq}. Since we do not have an analytical steady-state solution for the condensate wave functions and the cavity modes, we numerically diagonalize the Bogoliubov matrix for a numerically determined steady-state solution to obtain the collective excitation spectrum presented in~\fref{fig:coll_ex}.


%

\end{document}